\documentclass[showpacs,preprintnumbers,amsmath,amssymb]{revtex4}

\usepackage{graphicx}
\usepackage{dcolumn}
\usepackage{bm}
\usepackage{epsf}

\newcommand{\bea}{\begin{eqnarray}}
\newcommand{\eea}{\end{eqnarray}}
\newcommand{\beqa}{\begin{eqnarray}}
\newcommand{\eeqa}{\end{eqnarray}}
\newcommand{\eq}[1]{eq.~(\ref{#1})}
\newcommand{\eqs}[2]{eqs.(\ref{#1},\ref{#2})}
\newcommand{\eqss}[3]{eqs.(\ref{#1},\ref{#2},\ref{#3})}

\newcommand{\Eq}[1]{Eq.~(\ref{#1})}

\newcommand{\ur}[1]{(\ref{#1})}
\newcommand{\urs}[2]{(\ref{#1},\ref{#2})}
\newcommand{\urss}[3]{(\ref{#1},\ref{#2},\ref{#3})}

\newcommand{\beq}{\begin{equation}}
\newcommand{\eeq}{\end{equation}}
\newcommand{\la}[1]{\label{#1}}
\newcommand{\ba}{\begin{array}}
\newcommand{\ea}{\end{array}}
\newcommand{\ee}{\epsilon}

\newcommand{\half}{{\textstyle{\frac{1}{2}}}}
\newcommand{\at}{\overline{10}}

\newcommand{\noi}{\noindent}
\newcommand{\nn}{\nonumber}
\newcommand{\n}{\nonumber}

\newcommand{\octet}{$\left({\bf 8},\half^+\right)$}
\newcommand{\decuplet}{$\left({\bf 10},\frac{3}{2}^+\right)$}
\newcommand{\antidecuplet}{$\left({\bf\overline{10}},\half^+\right)$}

  \def\Tr{\mbox{Tr}}

 \def\Dirac#1{#1\hskip-6pt/}

\renewcommand{\d}{\dagger}

\newcommand{\scp}[2]{{\bf #1}\cdot{\bf #2}}

\renewcommand{\u}[2]{u_{#1}({\bf #2})}

\renewcommand{\v}[2]{v_{#1}({\bf #2})}

\newcommand{\inte}[1]{\int\!(d{\bf #1})}

\def\Dirac#1{#1\hskip-5pt/}

\renewcommand{\slash}[1]{\Dirac #1}

\begin{document}

\title{Estimate of the $\Theta^+$ width in the Relativistic Mean Field Approximation}
\author{Dmitri Diakonov$^{a,b}$}
\author{Victor Petrov$^a$}
\affiliation{ $^a$ St. Petersburg Nuclear Physics Institute,
Gatchina, 188 300, St. Petersburg, Russia\\
$^b$ NORDITA, Blegdamsvej 17, DK-2100 Copenhagen, Denmark
}

\date{June 8, 2005}

\begin{abstract}
In the Relativistic Mean Field Approximation three quarks in baryons from the
lowest octet and the decuplet are bound by the self-consistent chiral field,
and there are additional quark-antiquark pairs whose wave function also
follows from the mean field. We present a generating functional for the 3-quark,
5-quark, 7-quark ... wave functions inside the octet, decuplet and antidecuplet
baryons treated in a universal and compact way. The 3-quark components have the
$SU(6)$-symmetric wave functions but with specific relativistic corrections
which are generally not small. In particular, the normalization of the 5-quark
component in the nucleon is about 50\% of the 3-quark component. We give explicitly
the 5-quark wave functions of the nucleon and of the exotic $\Theta^+$. We develop
a formalism how to compute observables related to the 3- and 5-quark Fock components
of baryons, and apply it to estimate the $\Theta^+$ width which turns out to be
very small, 2-4 MeV, although with a large uncertainty.
\end{abstract}


\maketitle

\section{Introduction}

Were the chiral symmetry of the QCD lagrangian not broken spontaneously,
the nucleon would be either nearly massless or degenerate with its chiral partner,
$N(1535,\half^-)$. Both alternatives are many hundreds of MeV away from reality,
which serves as one of the most spectacular indications that chiral symmetry
is spontaneously broken. It also serves as a warning that if we disregard the
effects of the spontaneous chiral symmetry breaking we shall get nowhere in
understanding light baryons.

Spontaneous chiral symmetry breaking implies that at the microscopic level of QCD
nearly massless $u,d,s$ quarks gain a dynamical momentum-dependent mass $M(p)$
with $M_{u,d}(0)\approx 350\,{\rm MeV}$. A probable mechanism~\cite{DP86}
of how it happens is provided by instantons -- large fluctuations of the gluon field
in the vacuum. The resulting massive quarks are usually called the constituent
quarks; they necessarily, as a consequence of chiral symmetry, have to interact
with the (pseudo) Goldstone pion field, and actually very strongly: the dimensionless
coupling constant is about $M(0)/F_\pi\approx 4$. The corresponding low-energy
interaction lagrangian is written below, in Section II. It implies that inside
baryons there is a strong chiral field. Generally speaking, the chiral field
experiences quantum fluctuations; however, one may ask if it is reasonable to
introduce the notion of a mean chiral field inside baryons.

The mean field approach to bound states is usually justified by the large
number of participants. The Thomas--Fermi approximation to atoms is justified
at large $Z$, and the shell model for nuclei is justified at large $A$. In
baryons, the appropriate large parameter justifying the mean field approach
would be the number of colors $N_c$~\cite{Witten}. The number of colors being
$N_c\!=\!3$ in the real world, one may wonder how accurate is the mean-field
picture. Theoretically speaking, there are two kind of corrections in $1/N_c$
to the mean field.
One kind is due to the high-frequency fluctuations of the chiral field about
its mean-field value in a baryon. These are loop corrections and are additionally
suppressed by factors of $1/(2\pi)$. With the present precision, such corrections,
typically of the order of 10\%, can be ignored. The second type can be called
kinematical: they are due to the rotations of the baryon mean field in ordinary
and flavor spaces, and are not suppressed additionally. Such corrections are
not small at $N_c\!=\!3$ (although they tend to zero in the academic limit $N_c\to\infty$)
and should be taken into account exactly, if possible.

In this paper, we adopt the view~\cite{DP-CQSM} that there is a self-consistent
mean chiral field in baryons, which binds three massive constituent quarks, see Fig.~1.
The binding appears to be rather tight; bound-state quarks are relativistic
and their wave function has both the upper $s$-wave Dirac component and the
lower $p$-wave Dirac component, see Section III. Simultaneously, the negative-energy
Dirac sea of constituent quarks is distorted by the same mean field, leading to
the presence of an indefinite number of additional quark-antiquark ($\bar QQ$)
pairs in baryons, see Fig.~2. Ordinary baryons are superpositions of $3Q$, $5Q$, $7Q$...
Fock components. This picture which we shall call the Relativistic Mean Field Approximation
to baryons (or else the Chiral Quark Soliton Model where the word ``soliton " is an alias
of the mean field), leads, without any fitting parameters,
to a reasonable quantitative description of the baryons properties~\cite{DP-CQSM,Review},
including nucleon parton distributions at a low normalization point~\cite{SF} and
other baryon characteristics~\cite{GPV}. It should be stressed that the approximation
supports full relativistic invariance and all symmetries following from QCD.

We shall see that the normalization of the $5Q$ component in the nucleon is not
small as compared to its $3Q$ component. The three-quark picture of a nucleon is
an out-fashioned cartoon. It might do in popular lectures but professionals should
explain why the spin carried by three quarks is three times less, and the nucleon
$\sigma$-term is four times bigger than in the naive $3Q$ picture~\cite{D04}.
Taking into account the $\bar QQ$ pairs in the nucleon explains these paradoxes
\cite{DPPrasz,WY}.

\begin{figure}[htb]
\begin{minipage}[t]{.50\textwidth}
\includegraphics[width=0.75\textwidth]{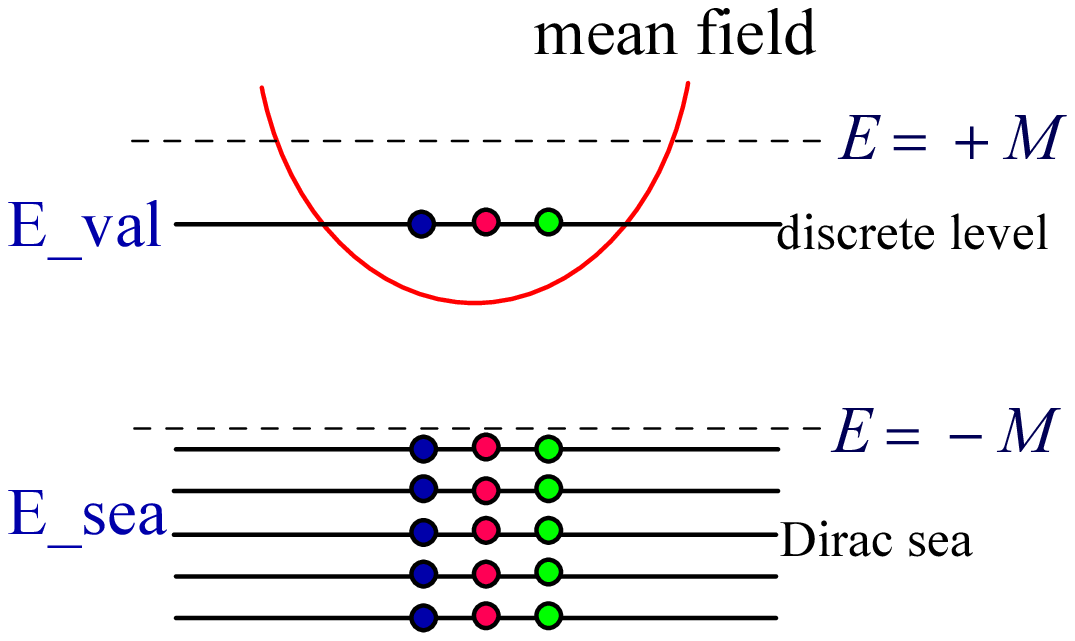}
\caption{A schematic view of baryons in the Relativistic Mean Field
Approximation. There are three ``valence'' quarks at a discrete
energy level created by the mean field, and the negative-energy
Dirac continuum distorted by the mean field, as compared to the free
one.} \label{fig:1}
\end{minipage}
\hfil
\begin{minipage}[t]{.45\textwidth}
\vspace{-4.2cm}
\includegraphics[width=0.58\textwidth]{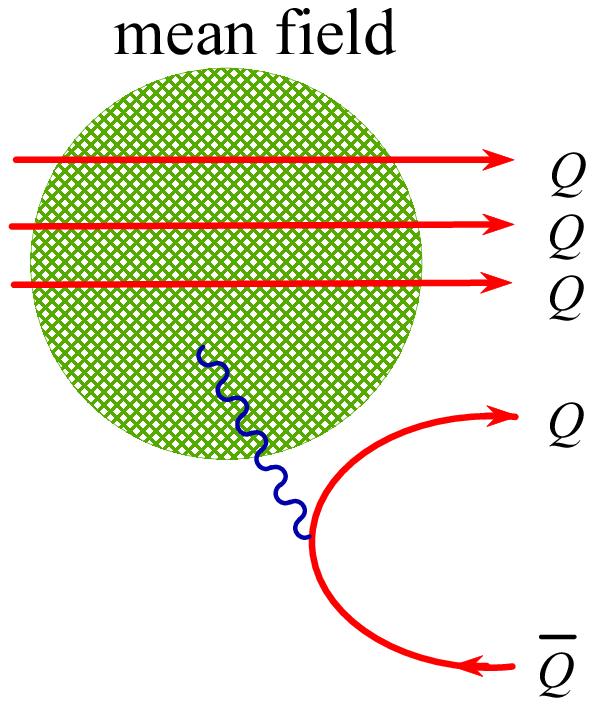}
\vspace{-0.5cm}
\caption{Equivalent view of baryons in the same approximation, where
the distorted Dirac sea is presented as $Q\bar Q$ pairs. The average number
of $Q\bar Q$ pairs is proportional to the amplitude squared of the mean field,
times $N_c$.} \label{fig:2}
\end{minipage}
\end{figure}

The correct lowest baryons' quantum numbers arise as a result of the quantization
of the rotation of the mean chiral field in the ordinary and in the flavor
spaces~\cite{Witten}. If the mean chiral field is presented as a unitary $3\times 3$
matrix $V(\vec x)$, the rotated field is
\beq
U(\vec x)=R\,V(\vec x)\,R^\dagger
\la{U}\eeq
where $R$ is an $SU(3)$ rotation matrix; it can be parameterized by eight ``Euler
angles" as it is done, for example, in Appendix A. We shall for simplicity set
the strange quark mass $m_s=0$; in this limit any rotated mean field \ur{U}
is, classically, as good as the un-rotated one: the baryon energy is degenerate
in rotations.

In quantum mechanics, however, the rotations are quantized. As first pointed out
by Witten~\cite{Witten} and then derived using different techniques by a number of
authors~\cite{quantization}, the quantization rule is such that the lowest baryon
multiplets are the octet with spin 1/2 and the decuplet with spin 3/2
({\it i.e.} exactly those observed in nature) followed by the exotic
{\it anti}-decuplet with spin 1/2 again. The parity of all rotational states is the same.
Those baryons are distinguished by the rotational wave functions depending on the
eight Euler angles parameterizing the $SU(3)$ rotation matrix $R$; the wave functions
are given explicitly in Section IV.

Qualitatively, one can think of different baryons as ``living'' in different
parts of an 8-dimensional globe parametrized by 8 ``Euler angles".
The $\Theta^+$ lives near the North pole of that globe, at least in the academic
limit of large $N_c$. The average polar angle for the rotational state corresponding
to the $\Theta^+$ vanishes as $1/\sqrt{N_c}$, see section IV.D. Therefore,
in the limit $N_c\to\infty$ one can approximate the rotation by small (kaon)
fluctuations about the North pole. Mathematically, it comes to the Callan--Klebanov
``bound-state approach to strangeness"~\cite{CK} where one studies the linear response
of a nucleon to a small-amplitude kaon perturbation, or the $KN$ scattering, to see if
there is a $\Theta^+$ resonance. In such approach the narrow $\Theta^+$ does
not exist, at least in the Skyrme model for the $KN$ scattering, unless one extends
the parameters of the model~\cite{KlebanovRho}. The Skyrme model for the
self-consistent chiral field is, however, not realistic, and it is unclear
what lesson can one draw from the existence or non-existence of a resonance
in this particular dynamical model.

Even more important, it is exactly the situation where the large $N_c$ limit
can hardly be trusted. In reality at $N_c=3$ the $\Theta^+$ rotational wave
function is spread over the whole 8-dimensional globe and is far from the
``North pole". A quantum-mechanical model of the situation has been suggested by Cohen~\cite{Cohen03}
and Pobylitsa~\cite{Pobylitsa03}; the model can be solved numerically at any
$N_c$~\cite{Cohen04}. It turns out that the energy levels at $N_c=3$ differ radically
from their positions at $N_c\to\infty$. Given this experience, we shall
treat the $\Theta^+$ rotational wave function {\it exactly} at $N_c=3$, see
section IV. At the same time we shall neglect the fluctuations of the chiral
field about its mean field value since these are suppressed additionally as are
any generic loop corrections.

In this approach, {\it all} low-energy properties of baryons from the
\octet, \decuplet {\it and} \antidecuplet multiplets (including {\it e.g.} parton
distributions at low virtuality) follow from the shape of the mean chiral field
in the common or `classical' baryon; the difference and splitting between baryons
from those multiplets arise exclusively from the difference in their rotational
wave functions. This difference can be immediately translated into the quark
wave functions of the individual baryons, both in the infinite
momentum~\cite{PP-IMF,DP-Fock} and the rest~\cite{D04-Minn} frames.
In Section III we present a compact general formalism how to find the 3-quark,
5-quark, 7-quark ... wave functions inside the octet, decuplet and antidecuplet
baryons, which is further detailed in Sections V and VI. In Section VII
we find the quark wave functions of the $3Q$ components in the octet and decuplet
baryons. In the non-relativistic limit (implying a weak mean field), we obtain
the old $SU(6)$ quark wave functions for the octet and decuplet baryons but
with well-defined relativistic corrections. The $5Q$ wave functions in the
ordinary and exotic baryons can be also found explicitly~\cite{D04-Minn,DP-Fock},
see Section VIII.

In Sections IX--XI we develop a formalism how to compute observables related to
the 3- and 5-quark Fock components of baryons, and apply it in Section XII to estimate
the nucleon axial constant and the transition matrix element of the strange axial current
between the $\Theta^+$ and the nucleon: it gives an estimate of the
$\Theta^+\to KN$ decay width. The latter turns out to be very small, 2-4 MeV,
although with a large uncertainty discussed in Section XIII.

The essence of QCD with its spontaneous breaking of chiral symmetry is that adding
a low-energy pseudoscalar meson (or a $\bar Q Q$ pair) to a baryon is equivalent
to rotating the vacuum state along the Goldstone valley, meaning no change of the
physical state. In order to separate the true $\bar QQ$ pairs in a baryon
from those in the vacuum, one has to consider baryons in the Infinite Momentum Frame
(IMF). In this and only this frame the true $\bar QQ$ pairs in a baryon have an
infinite momentum as contrasted to those in the vacuum, which have a finite momentum.
Therefore, an accurate definition of what are the 3-, 5-,... Fock components
of baryons can be made only in the IMF. It also has the advantage that the vector and
axial currents with a finite momentum transfer do not create or annihilate quarks with
infinite momenta. The baryon matrix elements are thus non-zero only between Fock components
with equal number of quarks and antiquarks.

Since the $\Theta^+$ has no $3Q$ component it means that one has to calculate the
matrix element between the $5Q$ component of the $\Theta^+$ and the $5Q$ component
of the nucleon. In principle, one has to add also the $7Q\to 7Q,\,9Q\to 9Q...$
transitions, but we neglect them in the present paper. To control this
approximation, we compute, using the same technique, the nucleon axial constant
$g_A(N)$. In the (very crude) non-relativistic $3Q$ approximation to nucleons, this
constant is approximately $5/3=1.667$; taking into account the $5Q$ component
of the nucleon moves it to the value of $1.36$ being already not too far from the
experimental value $g_A(N)=1.27$~\cite{PDG}. It should be noted that the summation of the
contributions of any number of $\bar QQ$ pairs in the Relativistic Mean Field
Approximation to nucleons moves $g_A(N)$ quite close to the experimental value~\cite{corrNc1}.
In the $5Q$ approximation to the $\Theta^+\to KN$ transition, we obtain
$g_A(\Theta\to KN)\approx 0.14-0.2$ leading to the estimate
$\Gamma_\Theta\approx 2-4\,{\rm MeV}$. In this estimate, we neglect the quark exchange
contributions to the $\Theta^+\to KN$ transition, which are potentially capable
of reducing further the width. Qualitatively, the axial constant of
the $\Theta^+\to KN$ transition is small because it is analogous not to the
large nucleon axial constant itself but to the {\it change} of this constant as
one goes from the $3Q$ to the $5Q$ contribution.

\section{The effective action}

The effective action approximating QCD at low momenta describes
``constituent'' quarks with the momentum dependent dynamical mass
$M(p)$ interacting with the scalar ($\Sigma$) and pseudoscalar
(${\bf \Pi}$) fields such that $\Sigma^2+{\bf \Pi}^2=1$ at spatial
infinity. The momentum dependence $M(p)$ serves as a formfactor of
the constituent quarks and provides the effective theory with the
ultraviolet cutoff. Simultaneously, it makes the theory non-local.
The action is~\cite{DP86}
\beq
\!S_{\rm eff}\!=\!\!\int\!\frac{d^4pd^4p'}{(2\pi)^8}\,
\bar\psi(p)\left[\slash{p}\,(2\pi)^4\delta^{(4)}(p-p')\!-\!\sqrt{M(p)}
\left(\Sigma(p\!-\!p')+i\Pi(p\!-\!p')\gamma_5\right)\sqrt{M(p')}\right]\psi(p'),
\la{action}\eeq
where $\psi,\bar\psi$ are quark fields carrying
color, flavor and Dirac bispinor indices. In the instanton model of
the QCD vacuum from where this action has been originally
derived the function $M(p)$ is such that there is no
real solution of the mass-shell equation $p^2=M(-p^2)$, therefore
quarks are not observable as asymptotic states, -- only their bound
states. However, this is not the true confinement. Unfortunately,
the instanton model's $M(p)$ has a cut at $p^2=0$ corresponding to
massless gluons left in that model. In the true confining theory
there should be no such cuts. Nevertheless, such $M(p)$ creates some
kind of a soft ``bag'' for quarks. Contrary to the naive bag
picture which does not respect relativistic invariance,
\eq{action} supports all general principles and sum rules for
conserved quantities.

The scalar, pseudoscalar~\cite{DP-preprint}, vector and axial~\cite{Bron-Dor}
mesons follow from the correlation functions computed from \eq{action}.
The light-cone quark wave functions of the pion and of the photon have
been found in Ref.~\cite{PPRWG}; the electromagnetic pion radius
has been computed in the original paper~\cite{DP86}.

Turning to baryons, the mean $\Sigma,\Pi$ field (called chiral field
for short in what follows) in the full non-local formulation
\ur{action} has been found by Broniowski, Golli and
Ripka~\cite{BGR}. It sets an example how one has to proceed in the
model calculations. However, to simplify the mathematics we shall
use here a more standard approach: we shall replace the constituent
quark mass by a constant $M=M(0)$ and mimic the decreasing function $M(p)$
by the UV Pauli--Villars cutoff~\cite{SF}.

\section{Baryon wave function in terms of quark creation-annihilation operators}

Let $a,a^\dagger({\bf p})$ and $b,b^\dagger({\bf p})$ be the
annihilation--creation operators of quarks and antiquarks
(respectively) of mass $M$, satisfying the usual anticommutator
algebra $\{a({\bf p})a^\d({\bf p'})\}=\{b({\bf p})b^\d({\bf p'})\}
=(2\pi)^3\delta^{(3)}({\bf p}-{\bf p'})$ and annihilating the vacuum state
$a,b|0\!\!>=\!0$, $<\!\!0|a^\d,b^\d\!=\!0$. For quarks, the
annihilation-creation operators carry, apart from the 3-momentum
${\bf p}$, also the color $\alpha$, flavor $f$ and spin $\sigma$
indices but we shall suppress them until they are explicitly needed.
The Dirac sea is presented by the coherent exponent of the quark and
antiquark creation operators~\cite{PP-IMF},
\beq
{\rm coherent\;exponent\;for}\;\bar QQ\;{\rm pairs}
=\exp\left(\inte{p}(d{\bf p'})\,a^\dagger({\bf p})\,
W({\bf p},{\bf p'})\,b^\dagger({\bf p'})\right)|0\!>,
\la{cohexp}\eeq
where $(d{\bf p})=d^3{\bf p}/(2\pi)^3$ and $W({\bf p_1},{\bf p_2})$
is the quark Green function at equal times in the background $\Sigma,{\bf \Pi}$
fields~\cite{PP-IMF,DP-Fock} (see Fig.~2); we shall specify the function $W$ below.
In the mean field approximation the chiral field is replaced by the spherically-symmetric
self-consistent field:
\beq
\pi({\bf x})={\bf n\cdot\tau}P(r),\qquad {\bf n}={\bf x}/r,\qquad
\Sigma({\bf x})=\Sigma(r).
\la{hh}\eeq
On the chiral circle (to which we restrict ourselves for simplicity)
$\Pi={\bf n\cdot\tau}\,\sin P(r),\;\Sigma(r)=\cos P(r)$
where $P(r)$ is the profile function of the self-consistent field.
It is fairly approximated by~\cite{DP-CQSM,DPPrasz}
\beq
P(r)=2\,{\rm atan}\left(\frac{r_0^2}{r^2}\right),\qquad r_0\approx \frac{0.8}{M},
\la{sc_prof}\eeq
where $M\approx 345\,{\rm MeV}$ is the dynamical quark mass at zero virtuality,
known to fit numerous observables within the instanton mechanism of the spontaneous
chiral symmetry breaking~\cite{DP86}.

The self-consistent chiral field \ur{sc_prof} creates a
bound-state level for quarks, whose wave function $\psi_{\rm lev}$
satisfies the static Dirac equation with eigenenergy $E_{\rm lev}$~\cite{KR,BB,DP-CQSM}:
\beq
\psi_{\rm lev}({\bf x})=\left(\ba{c}\ee^{ji}h(r)\\-i\ee^{jk}\,
({\bf\sigma\!\cdot\!n})^i_k\,j(r)
\ea\right),\qquad\left\{\ba{c}h'+h\,M\sin P-j(M\cos P+E_{\rm lev})=0,\\
j'+2j/r-j\,M\sin P-h(M\cos P-E_{\rm lev})=0,\ea\right.
\la{level}\eeq
where $i=1,2$ is the spin and $j=1,2=u,d$ is the isospin index.
In the non-relativistic limit ($E_{\rm lev}\approx M$) the $L\!=\!0$
upper component of the Dirac bispinor $h(r)$ is large while the $L\!=\!1$
lower component $j(r)$ is small. Solving \eq{level} for the self-consistent
field \ur{sc_prof} one finds that `valence' quarks are tightly bound
($E_{\rm lev}=200\,{\rm MeV}$) but the lower component $j(r)$ is
still substantially smaller than the upper one $h(r)$, see Figs.~3,4.

\begin{figure}[htb]
\begin{minipage}[t]{.45\textwidth}
\includegraphics[width=\textwidth]{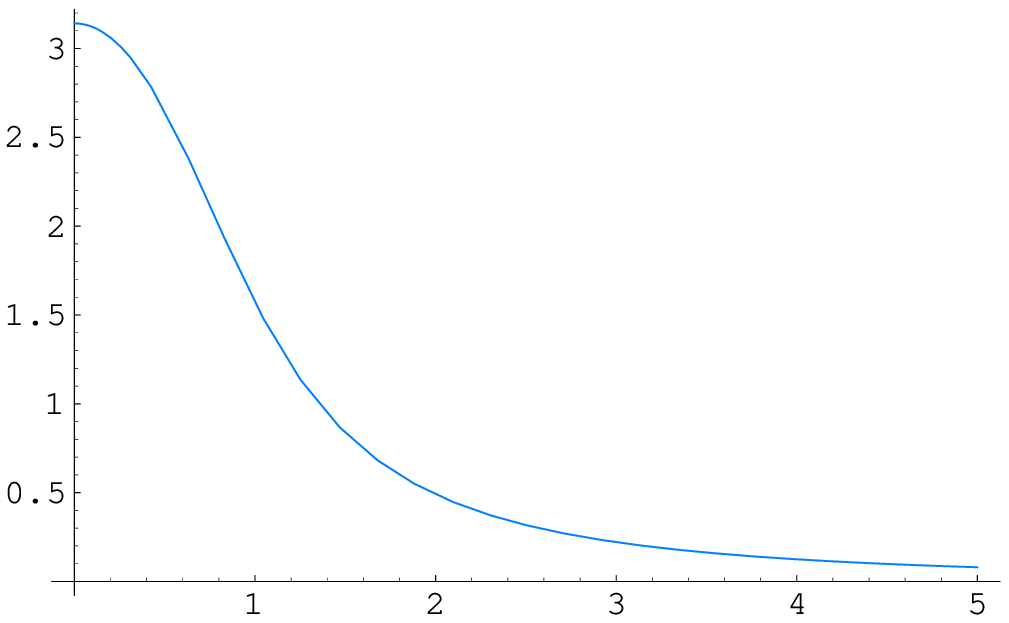}
\caption{The space profile of the self-consistent chiral field $P(r)$ in light baryons.
One unit on the horizontal axis is $r_0=0.8/M=0.46\,{\rm fm}$.} \label{fig:3}
\end{minipage}
\hfil
\begin{minipage}[t]{.45\textwidth}
\includegraphics[width=\textwidth]{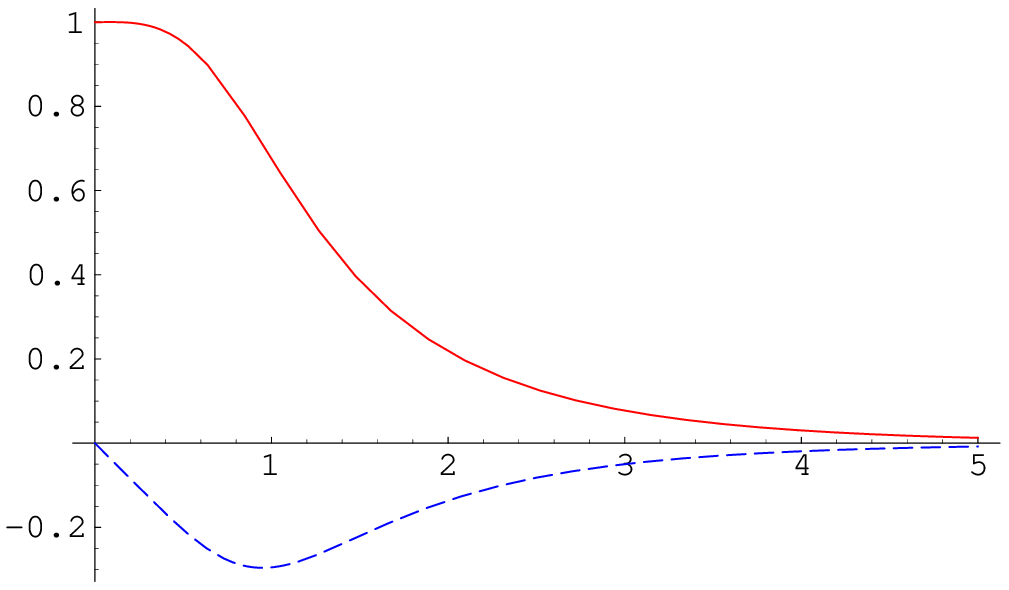}
\caption{Bound-state quark upper $s$-wave component $h(r)$ (solid) and the
lower $p$-wave component $j(r)$ (dashed) in light baryons. The three valence
quarks have the energy $E_{\rm lev}=200\,{\rm MeV}$ each.} \label{fig:4}
\end{minipage}
\end{figure}

The valence quark part of the baryon wave function is given by the
product of $N_c$ quark creation operators that fill in the discrete
level~\cite{PP-IMF}:
\beqa\la{val1}
{\rm valence\; quarks\; wave\; function}&=&\prod_{{\rm color}=1}^{N_c}
\int(d{\bf p})\,F({\bf p})\,a^\dagger({\bf p}),\\
F({\bf p})&=&\int\!(d{\bf p'})\sqrt{\frac{M}{\ee_p}}\!\left[\bar u({\bf p})
\,\gamma_0\,\psi_{\rm lev}({\bf p})\,(2\pi)^3\delta({\bf p}\!
-\!{\bf p'})\!-\!W({\bf p},{\bf p'})\,\bar v({\bf
p'})\,\gamma_0\,\psi_{\rm lev}(-\!{\bf p'})\right]\!,
\la{val2}\eeqa
where $\psi_{\rm lev}({\bf p})$ is the Fourier transform of
\eq{level}. The second term in \Eq{val2} is the contribution of the
distorted Dirac sea to the one-quark wave function. $u_\sigma({\bf p})$
and $v_\sigma({\bf p})$ are the plane-wave Dirac bispinors
projecting to the positive and negative frequencies, respectively.
In the standard basis they have the form
\beq\la{uv}
\u{\sigma}{p}=\left(\ba{c}\sqrt{\frac{\ee+M}{2M}}s_\sigma\\
\sqrt{\frac{\ee-M}{2M}}\frac{{\bf
p\cdot\sigma}}{|p|}s_\sigma\ea\right),\qquad
\v{\sigma}{p}=\left(\ba{c}\sqrt{\frac{\ee-M}{2M}}\frac{{\bf p\cdot\sigma}}{|p|}s_\sigma\\
\sqrt{\frac{\ee+M}{2M}}s_\sigma\ea\right),\qquad \bar u u=1=-\bar v v\,,
\eeq
where $\ee=\!+\!\sqrt{{\bf p}^2+M^2}$ and $s_\sigma$ are
two 2-component spinors normalized to unity, for example,
\beq
s_1=\left(\ba{c}1\\0\ea\right),\qquad
s_2=\left(\ba{c}0\\1\ea\right),\qquad \sigma=1,2.
\la{s}\eeq

The full baryon wave function is given by the product of the valence
part \ur{val1} and the coherent exponent \ur{cohexp} describing the
distorted Dirac sea. Symbolically, one writes the baryon wave
function in terms of the quark and antiquark creation operators
~\cite{PP-IMF}:
\beq B[a^\d,b^\d]=\prod_{{\rm color}=1}^{N_c}\int(d{\bf p})\,F({\bf p})\,a^\dagger({\bf p})\;
\exp\left(\inte{p}(d{\bf p'})\,a^\dagger({\bf p})\; W({\bf p},{\bf p'})\,
b^\dagger({\bf p'})\right)|0\!>.
\la{B1}\eeq

At this point one has to recall that the saddle point at the
self-consistent chiral field is degenerate in global translations
and global $SU(3)$ flavor rotations \ur{U} (the $SU(3)$ breaking by the
strange mass can be treated as a perturbation later). Integrating over
translations leads to the momentum conservation: the sum of all
quarks and antiquarks momenta have to be equal to the baryon
momentum. Integration over rotations $R$ leads to the projection of
the flavor state of all quarks and antiquarks onto the spin-flavor
state $B(R)$ describing a particular baryon from the $\left({\bf
8},\frac{1}{2}^+\right), \left({\bf 10},\frac{3}{2}^+\right)$ or
$\left({\bf \overline{10}},\frac{1}{2}^+\right)$ multiplet.

Restoring color ($\alpha=1,2,3$), flavor ($f=1,2,3$), isospin
($j=1,2$) and spin ($\sigma=1,2$) indices, the quark wave function
inside a particular baryon $B$ with spin projection $k$ is given, in
full glory, by \cite{PP-IMF,DP-Fock}
\beqa\nn
\Psi_k(B)&=&\int\!dR\,B^*_k(R)\,\ee^{\alpha_1\alpha_2\alpha_3}\prod_{n=1}^{3}
\int\!(d{\bf p_n})\,R^{f_n}_{j_n}\,F^{j_n\sigma_n}({\bf p_n})\,
a^\dagger_{\alpha_nf_n\sigma_n}({\bf p_n})\\
\la{Psi}
&\cdot &\exp\left(\int\!(d{\bf p})(d{\bf p'})\,
a^\dagger_{\alpha f\sigma}({\bf p})\,R^f_j\,
W^{j\sigma}_{j'\sigma'}({\bf p},{\bf p'})\,R^{\dagger\,j'}_{f'}\,
b^{\dagger\,\alpha f'\sigma'}({\bf p'})\right)|0\!\!>\,.
\eeqa
Acting on the vacuum state $|0\!>$ the operators $a^\dagger$ create three
`valence' quarks at the bound-state discrete level with the wave function $F$,
while the $a^\dagger,b^\dagger$ operators in the exponent create {\it any}
number of additional quark-antiquark pairs whose wave function is $W$.
\Eq{Psi} is thus a full relativistic field-theoretic description of
baryons, involving an infinite number of degrees of freedom.

Note that the three `valence' quarks are antisymmetric in color whereas the
additional $\bar Q Q$ pairs appear in color singlets. The spin-flavor quark
structure of a particular baryon arises from projecting a general
$QQQ+n\bar QQ$ state onto the quantum numbers of the baryon in question;
this is achieved by means of integrating over all spin-flavor rotations $R$
with the rotational wave function $B^*_k(R)$ unique for a given baryon.

The third row of the matrix $R^f_j,\,f=3,$ introduces strange quarks both
at the valence level and in the sea; hence hyperons with explicit strangeness
will, generally, have valence $s$ quarks, and non-strange baryons will
contain $\bar ss$ pairs, even though only the $u,d$ quarks are affected
by the chiral field \ur{hh}, which is reflected by the fact that the
valence-level wave function $F$ and the pair wave function $W$ have
not full $SU(3)$ but only isospin indices $j=1,2=u,d$.

\Eq{Psi} encodes an enormous amount of information as it is the generating
functional for the quark wave functions in {\it all} Fock components of baryons
from the lowest multiplets. Expanding the coherent exponent to the 0$^{\rm th}$,
1$^{\rm st}$, 2$^{\rm nd}$... order one reads off the 3-, 5-, 7-...
quark wave functions of a particular baryon from the octet, decuplet or antidecuplet.
All this information can be put in a compact form because the Relativistic Mean
Field Approximation is being used.

To make this powerful formula fully workable, we need to give
explicit expressions for the baryon rotational wave functions $B(R)$, the
valence wave function $F^{j\sigma}({\bf p})$ and the $\bar QQ$ wave
function in a baryon $W^{j\sigma}_{j'\sigma'}({\bf p},{\bf p'})$.

\section{Baryon rotational wave functions}

In general, baryon rotational states $B(R)$ are given by the $SU(3)$
Wigner finite-rotation matrices~\cite{hyperons}, and any particular
projection can be obtained by a routine $SU(3)$ Clebsch--Gordan
technique. However, in order to see the symmetries of the quark wave
functions it is helpful to use explicit expressions for $B(R)$, and
integrate over the Haar measure in \eq{Psi} explicitly.

We list below the rotational D-functions for the multiplets $\left({\bf 8},\frac{1}{2}\right)$,
$\left({\bf 10},\frac{3}{2}\right)$ and $\left({\bf\at},\frac{1}{2}\right)$
in terms of the product of the $R$ matrices. Since the projecting onto a specific
baryon in \eq{Psi} involves its conjugate rotational wave function, we list
the conjugate functions only. The un-conjugate ones are obtained by hermitian conjugation.

\subsection{$\left({\bf 8},\frac{1}{2}\right)$}

From the $SU(3)$ group point of view, the octet of baryons transforms exactly as an
octet of mesons; therefore, its rotational wave function can be composed of a
quark (transforming as $R$) and an antiquark (transforming as $R^\dagger$).
Accordingly, the rotational wave function of an octet baryon labeled by $a=1\ldots 8$
and having a spin index $k=1,2$ is
\beq
\left[D^{(8,\frac{1}{2})\,*}(R)\right]^a_k
\sim\,\ee_{kl}\,R^{\dagger\,l}_f\left(t^a\right)^f_g\,R^g_3.
\la{D8c}\eeq
where $\ee_{kl}$ is the antisymmetric $2\!\times\!2$ tensor and $t^a$ are the $SU(3)$
generators. In particular, the proton ($a=6+i7$) and neutron ($a=4+i5$) rotational
wave functions with spin $k=1,2$ are
\beq\la{DpDn}
p_k(R)^*=\sqrt{8}\,\ee_{kl}\,R^{\dagger\,l}_1\,R^3_3,\qquad
n_k(R)^*=\sqrt{8}\,\ee_{kl}\,R^{\dagger\,l}_2\,R^3_3.
\eeq

\subsection{$\left({\bf 10},\frac{3}{2}\right)$}

The decuplet states can be composed of three quarks; they are labeled by
a triple flavor index $\{f_1f_2f_3\}$ symmetrized in flavor and by a triple
spin index $\{k_1k_2k_3\}$ symmetrized in spin:
\beq
\left[D^{(10,\frac{3}{2})\,*}(R)\right]_{\{f_1f_2f_3\},\{k_1k_2k_3\}}
\sim\ee_{k'_1k_1}\ee_{k'_2k_2}\ee_{k'_3k_3}\,\left.R_{f_1}^{\dagger\,k'_1}\,R_{f_2}^{\dagger\,k'_2}\,
R_{f_3}^{\dagger\,k'_3}\right|_{{\rm sym\;in}\;\{f_1f_2f_3\}}\,.
\la{D10c}\eeq
For example, the $\Delta$-resonance rotational wave functions are
\beqa\la{DDeltappuu}
\Delta^{++},\;{\rm spin\;projection}\;+\!
\frac{3}{2}&:&\qquad\Delta^{++}_{\uparrow\uparrow}(R)^*=
\sqrt{10}\,R^{\dagger\,2}_1R^{\dagger\,2}_1R^{\dagger\,2}_1,\\
\la{Delta0u}
\Delta^{0},\,{\rm spin\;projection}\,+\!
\frac{1}{2}&:&\qquad \Delta^{0}_{\uparrow}(R)^*=
\sqrt{10}\,R^{\dagger\,2}_2(2R^{\dagger\,2}_1R^{\dagger\,1}_2\!
+\!R^{\dagger\,2}_2R^{\dagger\,1}_1)\,.
\eeqa

\subsection{$\left({\bf\at},\frac{1}{2}\right)$}

From the $SU(3)$ group point of view, the antidecuplet can be composed of three
antiquarks and its conjugate rotational wave function is
\beq
\left[D^{(\at,\frac{1}{2})\,*}(R)\right]_k^{\{f_1f_2f_3\}}\sim\left.
R^{f_1}_3\,R^{f_2}_3\,R^{f_3}_k\right|_{{\rm sym\;in}\;\{f_1f_2f_3\}}.
\la{Dat}\eeq
In particular,
\beqa
\la{DTheta}
\Theta^+,\;{\rm spin\;projection}\;k&:&\qquad \Theta_k(R)^*=\sqrt{30}\,R^3_3R^3_3R^3_k\,,\\
\la{DN*}
{\rm neutron}^*\;{\rm from}\;\at\;,\;{\rm spin\;projection}\;k&:&\qquad
n^{\at}_k(R)^*=\sqrt{10}\,(2R^2_3R^3_3R^3_k+R^3_3R^3_3R^2_k)\,.
\eeqa

All the rotational wave functions above are normalized in such a way that for any
(but the same) spin projection
\beq
\int\!dR\,B^*_{\rm spin}(R)\,B^{\rm spin}(R)=1;
\la{normB}\eeq
for different spin projections the integral is zero. Rotational wave functions
belonging to different baryons are also orthogonal. It can be easily checked
directly using the concrete parametrization of the $SU(3)$ rotation matrices
$R$ from Appendix A and performing the 8-dimensional integration with the
measure defined there.

\subsection{Large $N_c$ limit}

If $N_c$ is not equal to three but is treated as a free parameter, the
lightest baryons are not the octet, decuplet and antidecuplet but some
large $SU(3)$ multiplets whose dimensions depend on $N_c$. What $SU(3)$
multiplets are the large-$N_c$ prototypes of the usual multiplets at $N_c\!=\!3$,
is not uniquely defined. It seems natural to define the prototype multiplets
in such a way that their lightest members are ``nucleons" with spin
and isospin $\half$, ``$\Delta$'s" with spin and isospin $\frac{3}{2}$,
and ``$\Theta$" with spin $\half$ and isospin 0: this prescription is
sufficient to define unambiguously the large-$N_c$ prototypes of the octet,
decuplet and antidecuplet~\cite{DuPrasz,Cohen03,DP-03}.

The rotational wave functions of the large-$N_c$ analogs of the $N$,
$\Delta$ and $\Theta$ are obtained from \eqs{DpDn}{DDeltappuu} and
\ur{DTheta} by multiplying the corresponding equations by a factor
$\left(R^3_3\right)^{N_c-3}$. In Appendix A we give a concrete example
of the parametrization of a general $SU(3)$ rotation matrix $R$ in terms
of eight ``Euler'' angles. In fact they parameterize the $S^3\times S^5$
space, -- the direct product of the $3d$ and $5d$ spheres. In this
parametrization,
\beq
R^3_3=e^{i\alpha_{21}}\,\cos \phi_2\,\cos \theta,\qquad
\theta,\phi_2\in\left(0,\frac{\pi}{2}\right),
\la{R33}\eeq
where $\theta$ and $\phi_2$ can be viewed as polar angles of the $5d$ sphere.
It is clear that at $N_c\to\infty$ the rotational wave functions of the
``$N$", ``$\Delta$" and ``$\Theta$" are squeezed near the ``North pole" of
the sphere $S^5$ since the average polar angles vanish as $\theta,\phi_2
\sim 1/\sqrt{N_c}$. The rotated self-consistent field \ur{U} can be also
parameterized {\it {\`a} la} Callan--Klebanov~\cite{CK}:
\beq
U=RVR^\dagger=\sqrt{V}U_K\sqrt{V},\qquad
V({\bf x})=\left(\begin{array}{cc}
\exp \left[i(\mbox{\boldmath{$n$}}
\cdot\mbox{\boldmath{$\tau$}}) P(r)\right] &
\begin{array}{c} 0\\ 0\end{array}\\ \begin{array}{cc}0 &0 \end{array} & 1
\end{array}\right),
\la{CKpar}\eeq
where the meson $SU(3)$ unitary matrix $U_K$ is, for small meson fluctuations
$\phi$ about the self-consistent field $V$,
\beqa\la{UK}
&&U_K=1_3+i\phi^A\lambda^A, \qquad A=1...8,\\
&&\pi^{\pm}=\frac{\phi^1\pm i\phi^2}{\sqrt{2}},\quad\pi^0=\phi^3,\quad
K^+=\frac{\phi^4+i\phi^5}{\sqrt{2}},\quad
K^0=\frac{\phi^6+i\phi^7}{\sqrt{2}},\quad
\eta=\phi^8.
\la{mesons}\eeqa
One can compare both sides of \eq{CKpar} and find the meson fields in baryons
corresponding to rotations. In particular, for rotations ``near the North pole"
{\it i.e.} at small angles $\theta,\phi_2$, one finds the kaon field
\beq
K^+=-\sqrt{2}\,\sin\frac{P(r)}{2}
\left[\theta n_z+\phi_2(n_x-in_y)\right],\qquad
K^0=-\sqrt{2}\,\sin\frac{P(r)}{2}
\left[\theta(n_x+in_y)-\phi_2n_z\right],
\la{smallK}\eeq
meaning that at large $N_c$ the amplitude of the kaon fluctuations in the
prototype baryons ``$N$", ``$\Delta$" and ``$\Theta$" is vanishing as
$\sim 1/\sqrt{N_c}$. Therefore, the $\Theta$ problem becomes that of
the linear response of a nucleon to a small kaon fluctuation,
and can be studied in a particular model for the effective chiral
lagrangian~\cite{KlebanovRho}. However, in reality at $N_c\!=\!3$
the rotational wave functions of $N$ \ur{DpDn}, $\Delta$ \ur{DDeltappuu}
and $\Theta$ \ur{DTheta} correspond to large angles $\theta,\phi_2$
and are not concentrated near the ``North pole''. It means that
the kaon field in these baryons is generally not small. Therefore, in what
follows we shall use the exact $N_c\!=\!3$ rotational wave functions
\urss{DpDn}{DDeltappuu}{DTheta}.

\section{$\bar QQ$ pair wave function}

As explained in Refs.~\cite{PP-IMF,DP-Fock}, the pair wave function
$W^{j\sigma}_{j'\sigma'}({\bf p},{\bf p'})$ is expressed through the
finite-time quark Green function at equal times in the external
static chiral field \ur{hh}; schematically, it is shown in Fig.~2.
We shall need the Fourier transforms of the self-consistent chiral field,
\beq
\qquad
\Pi({\bf q})^j_{j'}=\int\!d^3{\bf x}\,e^{-i{\bf q\cdot x}}\,
({\bf n}\cdot\tau)^j_{j'}\,\sin P(r),\qquad
\Sigma({\bf q})^j_{j'}=\int\!d^3{\bf x}\,e^{-i{\bf q\cdot x}}\,
(\cos P(r)-1)\delta^j_{j'},
\la{SigmaPi_F}\eeq
where $\Pi({\bf q})$ is purely imaginary and odd while
$\Sigma({\bf q})$ is real and even.

In Refs.~\cite{PP-IMF,DP-Fock} a simplified interpolating approximation
for the pair wave function $W$ has been derived, which becomes exact in three
limiting cases: i) small pion field $P(r)$, ii) slowly varying $P(r)$,
iii) fast varying $P(r)$. In the infinite momentum frame the result is
a function of the fractions of the baryon's longitudinal momenta carried by
the quark ($z$) and antiquark ($z'$) of the pair, and their transverse momenta
${\bf p}_\perp,{\bf p'}_\perp$~\cite{footnote1}:
\beqa\nn
W^{j\sigma}_{j'\sigma'}(z,{\bf p}_\perp;z',{\bf p'}_\perp)&=&\frac{M{\cal M}}{2\pi Z}\,
\left\{\Sigma^{j}_{j'}({\bf q})\left[M(z'-z)\tau_3
+(z{\bf p'}-z'{\bf p})_\perp\cdot\tau_\perp\right]^{\sigma}_{\sigma'}\right.\\
\la{W}
&+&\left.i\,\Pi^{j}_{j'}({\bf q})\left[-M(z'+z){\bf 1}
+i\ee_{\alpha\beta}(z{\bf p'}-z'{\bf p})_{\perp\alpha}\tau_{\perp\beta}\right]^{\sigma}_{\sigma'}\right\},\\
\nn
Z&= &{\cal M}^2zz'(z+z')+z(p^{'2}_\perp+M^2)+z'(p^{2}_\perp+M^2),\qquad
{\bf q}=\left(({\bf p}+{\bf p'})_\perp, (z+z'){\cal M}\right).
\eeqa
\begin{figure}[htb]
\begin{minipage}[t]{\textwidth}
\includegraphics[width=0.3\textwidth]{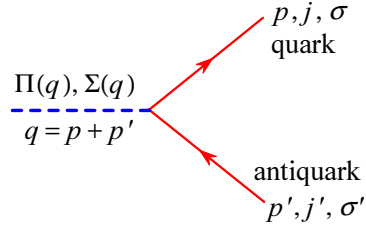}
\caption{The $\bar QQ$ pair wave function in a baryon in the Relativistic Mean Field
Approximation is related to the Fourier transform of the static self-consistent
chiral field.}
\label{fig:5}
\end{minipage}
\end{figure}
Here $j,j'=1,2$ are the isospin and $\sigma,\sigma'=1,2$ are the spin projections,
$\tau_{1,2,3}$ are Pauli matrices, $\ee_{\alpha\beta}$ is the antisymmetric $2d$ tensor,
the primed indices refer to the antiquark; ${\cal M}$ is the baryon and $M$ is the
constituent quark masses, ${\bf q}$ is the 3-momentum of the pair as a whole, transferred
from the background field $\Sigma({\bf q}),\Pi({\bf q})$.

The pair wave function $W$ is normalized in such a way that the creation-annihilation
operators in \eq{Psi} satisfy the anticommutation relations
\beq
\{a^{\alpha_1f_1\sigma_1}(z_1,{\bf p}_{1\perp}),
a^\d_{\alpha_2f_2\sigma_2}(z_2,{\bf p}_{2\perp})\}=
\delta^{\alpha_1}_{\alpha_2}\delta^{f_1}_{f_2}\delta^{\sigma_1}_{\sigma_2}\,
\delta(z_1-z_2)\,(2\pi)^2\delta^{(2)}({\bf p}_{1\perp}-{\bf p}_{2\perp})
\la{antcIMF}\eeq
and similarly for $b,b^\d$, and the integrals over momenta there are understood as
$\int\!dz\!\int d^2{\bf p}_\perp/(2\pi)^2$.

The pair wave function can be written in a more compact form by introducing the
fraction of the longitudinal momentum of the pair carried by the antiquark $y$,
and the transverse combination ${\bf Q_\perp}$,
\beq
y=\frac{z'}{z+z'},\qquad q_z=\frac{z+z'}{{\cal M}},\qquad
{\bf Q}_\perp=\frac{z{\bf p'}_\perp-z'{\bf p}_\perp}{z+z'}\,.
\la{zamena}\eeq
With this substitution \eq{W} takes the form
\beq
W^{j\sigma}_{j'\sigma'}(y,{\bf q},{\bf Q}_\perp)=\frac{M{\cal M}}{2\pi}\,
\frac{\Sigma^{j}_{j'}({\bf q})\left[M(2y-1)\tau_3
+{\bf Q}_\perp\cdot\tau_\perp\right]^{\sigma}_{\sigma'}
+i\,\Pi^{j}_{j'}({\bf q})\left[-M{\bf 1}
+i\ee_{\alpha\beta}{\bf Q}_{\perp\alpha}\tau_{\perp\beta}\right]^{\sigma}_{\sigma'}}
{{\bf Q}_\perp^2+M^2+y(1-y){\bf q}^2}\,.
\la{W1}\eeq

\section{Bound-state wave function}

As seen from \eq{val2}, the discrete-level wave function
$F^{j\sigma}({\bf p}) =F^{j\sigma}_{\rm lev}({\bf
p})+F^{j\sigma}_{\rm sea}({\bf p})$ consists of two pieces: one is
directly the wave function of the valence level, the other is
related to the change of the number of quarks at the discrete level
as due to the presence of the Dirac sea; it is a relativistic effect
and can be ignored in the non-relativistic limit, together with the
lower $L\!=\!1$ component $j(r)$ of the level wave function. Indeed,
in the baryon rest frame the evaluation of the first term in
\eq{val2} gives
\beq
F^{j\sigma}_{\rm lev}({\bf p})=\ee^{j\sigma}
\left(\sqrt{\frac{E_{\rm lev}+M}{2E_{\rm
lev}}}h(p)+\sqrt{\frac{E_{\rm lev}-M}{2E_{\rm lev}}}j(p)\right),
\la{Flev_rest}\eeq
where $h(p),j(p)$ are the Fourier transforms of the
valence wave functions \ur{level}:
\beqa\la{Fh}
h(p)&=&\int\!d^3x\,e^{-i\scp{p}{x}}\,h(r)
=4\pi\int\!dr\,r^2\frac{\sin\,pr}{pr}\,h(r),\\
\la{Fj}
j^a(p)&=&\int\!d^3x\,e^{-i\scp{p}{x}}\,(-in^a)j(r)
=\frac{p^a}{|p|}j(p),\qquad
j(p)=\frac{4\pi}{p^2}\!\int\!dr\,(pr\,\cos\,p r-\sin\,p r)\,j(r).
\eeqa
One sees that the second term in \eq{Flev_rest} is
double-suppressed in the non-relativistic limit $E_{\rm lev}\approx
M$: first, owing to the kinematical factor, second, since in this
limit the $L\!=\!1$ wave $j(r)$ is much less than the $L\!=\!0$ wave
$h(r)$.

In the infinite momentum frame the evaluation of the bispinors $\bar
u,\bar v$ from \eq{uv} produces~\cite{PP-IMF,DP-Fock}
\beq
F^{j\sigma}_{\rm lev}(z,p_\perp)=\sqrt{\frac{{\cal M}}{2\pi}}
\left[\ee^{j\sigma}h(p)+\left(p_z{\bf 1}+i\ee_{\alpha\beta}p_{\perp\alpha}
\tau_{\perp\beta}\right)^{\sigma}_{\sigma'}\ee^{j\sigma'}\frac{j(p)}{|p|}
\right]_{p_z=z{\cal M}-E_{\rm lev}}
\la{Flev_IMF}\eeq

Similarly, the evaluation of the ``sea'' part of the discrete-level
wave function gives
\beq
F^{j\sigma}_{\rm sea}(z,p_\perp)=-\sqrt{\frac{{\cal M}}{2\pi}}
\!\int\!dz'\frac{d^2p'_\perp}{(2\pi)^2}\,W^{j\sigma}_{j'\sigma'}(p,p')\,\ee^{j'\sigma''}\,
\left[(\tau_3)^{\sigma'}_{\sigma''}h(p')
-(\tau\cdot{\bf p'})^{\sigma'}_{\sigma''}\frac{j(p')}{|p'|}\right]_{p_z=z{\cal M}-E_{\rm lev}}
\la{Fsea_IMF}\eeq
where the pair wave function \ur{W} has to be used.

In the following evaluation of the baryon matrix elements we shall neglect the
relativistic effects in the discrete level wave function replacing it by the
first term in \eq{Flev_IMF}:
\beq
F^{j\sigma}(z,p_\perp)\approx \sqrt{\frac{{\cal M}}{2\pi}}
\left.\ee^{j\sigma}h(p)\right|_{p_z=z{\cal M}-E_{\rm lev}}.
\la{F_approx}\eeq

We have now fully determined all quantities entering the master
\eq{Psi} for the 3,5,7... Fock components of baryons' wave
functions.

\section{3-quark components of baryons}

The absolute majority of baryon models focus on the 3-quark Fock components of
the usual (non-exotic) baryons. We have already mentioned in the Introduction
that it is a crude approximation to reality: the 5-, 7-,... quark components
in the nucleon are not only non-negligible but critical for explaining such
important characteristics as the nucleon $\sigma$ term or the fraction of
nucleon spin carried by quarks. Nevertheless, the 3-quark component is
definitely important. In this section we derive the $3Q$ wave functions of
the octet and decuplet baryons from our master equation \ur{Psi} and show
that in the non-relativistic limit they become the well-known $SU(6)$ wave
functions of the old constituent quark model.

One gets the $3Q$ component of a baryon by ignoring the coherent exponent
with $Q\bar Q$ pairs in \eq{Psi}; each of the three valence quarks is rotated by
the matrix $R^f_j$ where $f=1,2,3$ is the flavor and $j=1,2$ is the isospin
index. To obtain the color-flavor-spin-space $3Q$ wave function of a particular
baryon from the \octet or the \decuplet, one has to integrate in \eq{Psi} over
all 8-parameter $SU(3)$ rotations $R$ with the (conjugate) rotational wave
function $B^*_k(R)$ corresponding to the chosen baryon with spin projection $k$.
These functions are given in Section IV. The arising $SU(3)$ group integrals are of
the type
\beq
T(B)^{f_1f_2f_3}_{j_1j_2j_3,k}
\equiv\int\!dR\,B^*_{k}(R)\,R^{f_1}_{j_1}R^{f_2}_{j_2}R^{f_3}_{j_3}
\la{T3}\eeq
where the three unitary matrices $R^{f_1}_{j_1},R^{f_2}_{j_2},R^{f_3}_{j_3}$
rotate the flavor of the quarks on the discrete level.
These tensors are computed in Appendix B: for baryons from the \octet
the relevant integral is \eq{4+1}, and for the \decuplet it is \eq{3+3}.
The tensor $T$ must be now contracted with the three discrete-level wave
functions from Section VI
\beq
F^{j_1\sigma_1}(p_1)F^{j_2\sigma_2}(p_2)F^{j_3\sigma_3}(p_3).
\la{3F}\eeq

In general the $3Q$ wave function depends on all four quark ``coordinates":
the position in space (${\bf r}$) (or the three-momentum ${\bf p}$),
the color ($\alpha$), the flavor ($f$) and the spin ($\sigma$), and also
on the baryon spin projection $k$. The wave function must be antisymmetric
under permutation of all four ``coordinates" for a pair of quarks. We suppress
the trivial color wave function $\ee^{\alpha_1\alpha_2\alpha_3}$ which factors
out. In the non-relativistic approximation we use the simplified level wave function
\ur{F_approx} and for clarity pass back to the coordinate space. We thus obtain,
for example, the {\bf $3Q$ wave function of the neutron}:
\beqa\nn
\left(|n\!>_k\right)^{f_1f_2f_3,\sigma_1\sigma_2\sigma_3}
({\bf r_1,r_2,r_3}) &=& \frac{\sqrt{8}}{24}\,
\ee^{f_1f_2}\,\ee^{\sigma_1\sigma_2}\,\delta^{f_3}_2\,\delta^{\sigma_3}_k\,
h(r_1)h(r_2)h(r_3)\\
&+&{\rm permutations\;of\;1,2,3},
\la{n1}\eeqa
times the antisymmetric $\ee^{\alpha_1\alpha_2\alpha_3}$ in color. In this
equation the flavor indices assume only two values: $f_{1,2,3}=1,2=u,d$.
\Eq{n1} says that the neutron spin is carried by the $d$-quark, and the
$ud$ pair is in the spin- and isospin-zero combination. It is better known in the form
\beqa\nn
|n\!\uparrow>&=&2\,d\!\uparrow\!(r_1)d\!\uparrow\!(r_2)
u\!\downarrow\!(r_3)\!-\!d\!\uparrow\!(r_1)u\!\uparrow\!(r_2)
d\!\downarrow\!(r_3)\!-\!u\!\uparrow\!(r_1)d\!\downarrow\!(r_2)
d\!\uparrow\!(r_3)\\
&+& {\rm permutations\;of\;} r_1,r_2,r_3,
\la{n2}\eeqa
which is the well-known non-relativistic $SU(6)$ wave function of the nucleon,
with a concrete space distribution of quarks, shown in Fig.~4.

Similarly, the {\bf $3Q$ wave function of the $\Delta^0$ resonance} with spin
projection 1/2, which may be compared with that of the
neutron, can obtained from the group integral \ur{3+3}, and reads
\beqa \nn
|\Delta^0\!\uparrow>^{f_1f_2f_3,\sigma_1\sigma_2\sigma_3}
({\bf r_1,r_2,r_3})&=&\frac{\sqrt{10}}{30}\,
\left(\delta^{f_1}_1\delta^{f_2}_2\delta^{f_3}_2
+\delta^{f_1}_2\delta^{f_2}_1\delta^{f_3}_2
+\delta^{f_1}_2\delta^{f_2}_2\delta^{f_3}_1\right)\\
\la{Delta01}
&&\cdot\left(\delta^{\sigma_1}_1\delta^{\sigma_2}_1\delta^{\sigma_3}_2
+\delta^{\sigma_1}_1\delta^{\sigma_2}_2\delta^{\sigma_3}_1
+\delta^{\sigma_1}_2\delta^{\sigma_2}_1\delta^{\sigma_3}_1\right)
h(r_1)h(r_2)h(r_3)
\eeqa
which can be also presented as a familiar $SU(6)$ wave function
\beqa \nn
|\Delta^0\!\uparrow>&=&
u\!\uparrow\!(r_1)\,d\!\uparrow\!(r_2)\,d\!\downarrow\!(r_3)+
d\!\downarrow\!(r_1)\,u\!\uparrow\!(r_2)\,d\!\uparrow\!(r_3)+
d\!\uparrow\!(r_1)\,d\!\uparrow\!(r_2)\,u\!\downarrow\!(r_3)\\
&+&{\rm permutations\;of\;} r_1,r_2,r_3.
\la{Delta02}\eeqa
There are, of course, relativistic corrections to these
$SU(6)$-symmetric formulae, arising from i) exact treatment of the
discrete level, \eqs{Flev_IMF}{Fsea_IMF}, and ii) additional $Q\bar
Q$ pairs described by \eq{W}. Both effects are generally not small.

\section{5-quark components of baryons}

One gets the wave functions of the $5Q$ component of baryons by expanding
the coherent exponent in the generating functional \ur{Psi} to the linear order
in the $\bar QQ$ pair. The $SU(3)$ group integral involves now three $R$'s
from the level and $R,R^\dagger$ from the pair, times the (conjugate)
rotational wave function $B^*_k(R)$ of the baryon in question:
\beq
T(B)^{f_1f_2f_3f_4,j_5}_{j_1j_2j_3j_4,f_5,k}
=\int\!dR\,B^*_{k}(R)\,R^{f_1}_{j_1}R^{f_2}_{j_2}R^{f_3}_{j_3}
R^{f_4}_{j_4}R^{\d\,j_5}_{f_5}.
\la{T5}\eeq
We shall systematically attribute the indices 1,2,3 to the valence quarks,
index 4 to the extra quark of the $\bar QQ$ pair, and index 5 to the antiquark.
The group integral \ur{T5} is computed in Appendix B: for octet baryons
the result is given in \eq{5+2} and for the antidecuplet baryons it is given in
\eq{7+1A}. To obtain the $5Q$ wave function of a baryon, one has to contract
$T$ from \eq{T5} with three valence quark wave functions $F$ \ur{3F} and with
the pair wave function $W$ \ur{W}.

In general, the $5Q$ wave functions look rather complicated as they depend
on five quark ``coordinates'', including their coordinates proper (or 3-momenta),
spin, flavor and color. We do not write explicitly the color degrees of freedom
but always imply that the $(1,2,3)$ quarks of the level are
antisymmetric in color while the quark-antiquark pair $(4,5)$ is a
color singlet, as it follows from \eq{Psi}. For example, the {\bf $5Q$
wave function of the neutron} is
\beqa \nn
\left(|n\!>_k\right)^{f_1f_2f_3f_4,\sigma_1\sigma_2\sigma_3\sigma_4}_{f_5,\sigma_5}
({\bf p_1\ldots p_5})&=&\frac{\sqrt{8}}{360}\,
F^{j_1\sigma_1}({\bf p_1})F^{j_2\sigma_2}({\bf p_2})F^{j_3\sigma_3}({\bf p_3})
W^{j_4\sigma_4}_{j_5\sigma_5}({\bf p_4,p_5})\\
\nn &\cdot &
\!\left\{\ee^{f_1f_2}\epsilon_{j_1j_2}\left[\delta^{f_3}_2\delta^{f_4}_{f_5}
\left(4\delta^{j_5}_{j_4}\delta^{k^\prime}_{j_3}
-\delta^{j_5}_{j_3}\delta^{k^\prime}_{j_4}\right)+
\delta^{f_4}_2\delta^{f_3}_{f_5}\left(4\delta^{j_5}_{j_3}\delta^{k^\prime}_{j_4}
-\delta^{j_5}_{j_4}\delta^{k^\prime}_{j_3}\right)\right]\right.\\
\nn &+&
\!\!\left.\ee^{f_1f_4}\epsilon_{j_1j_4}\left[\delta^{f_2}_2\delta^{f_3}_{f_5}
\left(4\delta^{j_5}_{j_3}\delta^{k^\prime}_{j_2}\!
-\!\delta^{j_5}_{j_2}\delta^{k^\prime}_{j_3}\right)+
\delta^{f_3}_2\delta^{f_2}_{f_5}
\left(4\delta^{j_5}_{j_2}\delta^{k^\prime}_{j_3}\!
-\!\delta^{j_5}_{j_3}\delta^{k^\prime}_{j_2}\right)\right]\right\}
\epsilon_{k^\prime k}
\\
&+&\!{\rm permutations\; of}\; (1,2,3).
\la{n5}\eeqa
Terms of the type of $\delta^{f_3}_{f_5}$ mean the flavor-symmetric combination
$s\bar s+u\bar u+d\bar d$, however quarks from this combination are partly
inside the pair wave function $W$ but partly in the ``valence'' bound state.
We have not invented how to present it in a more
compact form; however, \eq{n5} is a working formula which allows to get compact
physical results, see Section XII. The {\bf $5Q$ wave function of the proton}
is the same, with the replacement $\delta_2^{f_{1,2,3,4}}\to\delta_1^{f_{1,2,3,4}}$,
meaning that one $d$-quark must be replaced by the $u$-quark.\\

Turning to the exotic baryons from the $\left({\bf \at},\half^+\right)$,
projecting the three quarks from the discreet level onto the antidecuplet
rotational function \ur{Dat} gives an identical zero in accordance with
the fact that the exotic antiducuplet cannot be made of 3 quarks, see
\eq{6A}. The non-zero projection is achieved when one expands the coherent
exponent at least to the linear order. For example, one gets then from
\eqs{DTheta}{7+1Theta} the {\bf $5Q$ wave function of the $\Theta^+$}:
\beqa\nn
|\Theta^+_k\!>^{f_1f_2f_3f_4,\sigma_1\sigma_2\sigma_3\sigma_4}_{f_5,\sigma_5}
({\bf p_1\ldots p_5})&=&\frac{\sqrt{30}}{180}\,
\ee^{f_1f_2}\ee^{f_3f_4}\delta^3_{f_5}\,\ee_{j_1j_2}\ee_{j_3j_4}\,
F^{j_1\sigma_1}({\bf p_1})F^{j_2\sigma_2}({\bf p_2})F^{j_3\sigma_3}({\bf p_3})
W^{j_4\sigma_4}_{j_5\sigma_5}({\bf p_4,p_5})\\
&+&{\rm permutations\;of\;(1,2,3)}.
\la{Theta1}\eeqa
The color structure of the antidecuplet wave function is
$\ee^{\alpha_1\alpha_2\alpha_3}\delta^{\alpha_4}_{\alpha_5}$.
The quark flavor indices are $f_{1\!-\!4}=1,2=u,d$, and the antiquark
is $\bar s$ owing to $\delta^3_{f_5}$. Naturally, we have obtained
the quark content $\Theta^+=uudd\bar s$ where the two $(ud)$ pairs are
in the isospin-zero combination, thanks to $\ee^{f_1f_2}\ee^{f_3f_4}$.

To make contact with other work where the $\Theta^+$ wave functions
were obtained in various non-relativistic models or discussed in that
framework~\cite{nrm}, one has to pass to the coordinate space and
write \eq{Theta1} in the $\Theta^+$ rest frame using the non-relativistic
approximation \ur{F_approx} for the level wave function. We obtain
\beqa\nn
|\Theta^+_k\!>^{f_1f_2f_3f_4,\sigma_1\sigma_2\sigma_3\sigma_4}_{f_5,\sigma_5}
({\bf r_1\ldots r_5})
&=&\frac{\sqrt{30}}{180}\,\ee^{f_1f_2}\ee^{f_3f_4}\delta^3_{f_5}\,\ee^{\sigma_1\sigma_2}\,
h(r_1)h(r_2)h(r_3)\,W^{\sigma_3\sigma_4}_{k\,\sigma_5}({\bf r_4,r_5})\\
&+&{\rm permutations\;of\;(1,2,3)}
\la{Theta2}\eeqa
where the pair wave function in the coordinate space $W({\bf r_4,r_5})$
can be found in Ref.~\cite{D04-Minn}. The structure
$\ee^{f_1f_2}\ee^{\sigma_1\sigma_2}$ clearly shows that there is a
pair of $ud$ quarks in the spin and isospin zero combination,
exactly as in the nucleon, \eq{n1}. However, it does not mean that
there are prominent scalar isoscalar diquarks either in the nucleon
or in the $\Theta^+$: that would require their spatial correlation
which, as we see, is absent in the mean field approximation.
The $Q\bar Q$ pair wave function $W$ is a combination of four
partial waves with different permutation symmetries, in accordance with
the general analysis by Bijker, Giannini and Santopinto, Ref.~\cite{nrm}.
The amplitudes of those partial waves depend separately on the coordinates
${\bf r_{4,5}}$ measured from the baryon center of mass. More explicit
formulae are given in Ref.~\cite{D04-Minn}.

\section{Three quarks: normalization, vector and axial charges}

The normalization of a baryon wave function in the
second-quantization representation \ur{Psi} is found from
\beq
{\cal N}(B)=\frac{1}{2}\delta^k_l\,<\!\Psi^{\d\,B\,l}\Psi^B_k\!>.
\la{N1}\eeq
The annihilation operators in $\Psi^{\d\,B\,l}$ must be
dragged to the right where they ultimately nullify the vacuum state
$|0\!\!>$ and the creation operators from $\Psi^B_k$ should be
dragged to the left where they ultimately nullify the vacuum state
$<\!\!0|$. The result is non-zero owing to the anticommutation
relations \ur{antcIMF} or the ``contractions'' of the operators.

For the $3Q$ Fock component of a baryon, there are $3!$ possible
(and equivalent) contractions, and the ensuing contraction in color
indices gives another factor of
$3!=\ee^{\alpha_1\alpha_2\alpha_3}\ee_{\alpha_1\alpha_2\alpha_3}$.
Flavor projecting to a baryon with specific quantum numbers gives the
tensor \ur{T3}, or its hermitian conjugate for the conjugate wave
function. Hence the normalization of the $3Q$ component, shown schematically
in Fig.~6, left, is
\beqa\nn
{\cal N}^{(3)}(B)&=&\frac{(6\!\cdot\!6)\!}{2}\,\delta^k_l\,
T(B)^{f_1f_2f_3}_{j_1j_2j_3,k}\,T(B)^{l_1l_2l_3,l}_{f_1f_2f_3}
\int\!dz_{1,2,3}\!\int\frac{d^2p_{1,2,3}}{(2\pi)^6}\,\delta(z_1+z_2+z_3-1)\\
&&\cdot (2\pi)^2 \delta((p_1\!+\!p_2\!+\!p_3)_\perp)\,
F^{j_1\sigma_1}(p_1)F^{j_2\sigma_2}(p_2)F^{j_3\sigma_3}(p_3)
\,F^\d_{l_1\sigma_1}(p_1)F^\d_{l_2\sigma_2}(p_2)F^\d_{l_3\sigma_3}(p_3)
\la{N31}\eeqa
where $F^{j\sigma}(z,p_{\perp})$ are the level wave
functions \urs{Flev_IMF}{Fsea_IMF}. In the non-relativistic limit
$F^{j\sigma}(p)F^\d_{l\sigma}(p)\sim\delta^j_l\,h^2(p)$, see \eq{F_approx}.
Therefore in this simple case the normalization is the full contraction of the
two $T$ tensors, times an integral over momenta which can be
performed numerically once the level wave function $h(p)$ is known.

A typical physical observable is the matrix element of some operator
(which should be written down in terms of the quark
annihilation-creation operators $a,b,a^\d,b^\d$) sandwiched between
the initial and final baryon wave functions \ur{Psi}. We shall consider
as examples the operators of the vector and axial charges which can
be written through the annihilation-creation operators as
\beqa\nn
\left\{\ba{c}Q\\Q_5\ea\right\}&=&\int\!d^3x\,\bar\psi_e\,J^e_h\,
\left\{ \ba{c} \gamma_0 \\
\gamma_0\gamma_5 \ea \right\} \psi^h
=\int\!dz\,\frac{d^2p_\perp}{(2\pi)^2} \left[
a^\d_{e\pi}(z,p_\perp)a^{h\rho}(z,p_\perp)\,J^e_h
\left\{\ba{c}\delta^{\pi}_{\rho}\\(-\sigma_3)^{\pi}_{\rho}\ea\right\}
\right.
\\
&-&\left. b^{\d\,h\rho}(z,p_\perp)b_{e\pi}(z,p_\perp)\,J^e_h
\left\{\ba{c}\delta^{\pi}_{\rho}\\
(-\sigma_3)^{\pi}_{\rho}\ea\right\}\right]
\la{vac}\eeqa
where $J^e_h$ is the flavor content of the charge, and $\pi,\rho=1,2$
are helicity states. For example, if we
consider the $\rho^+=\bar d u$ current which annihilates $u$ quarks
and creates $d$ quarks and annihilates $\bar d$ antiquarks and
creates $\bar u$ quarks, the flavor currents in \eq{vac} are
$J^e_h(\rho^+)=\delta^e_2\delta^1_h$. Notice that there are no $a^\d
b^\d$ or $ab$ terms in the charges. This is a great advantage of the
infinite momentum frame where the number of $Q\bar Q$ pairs is not changed by the
current. Hence there will be only diagonal transitions between Fock
components with equal numbers of pairs, see Fig.~6, right.

\begin{figure}[htb]
\begin{minipage}[t]{.45\textwidth}
\includegraphics[width=\textwidth]{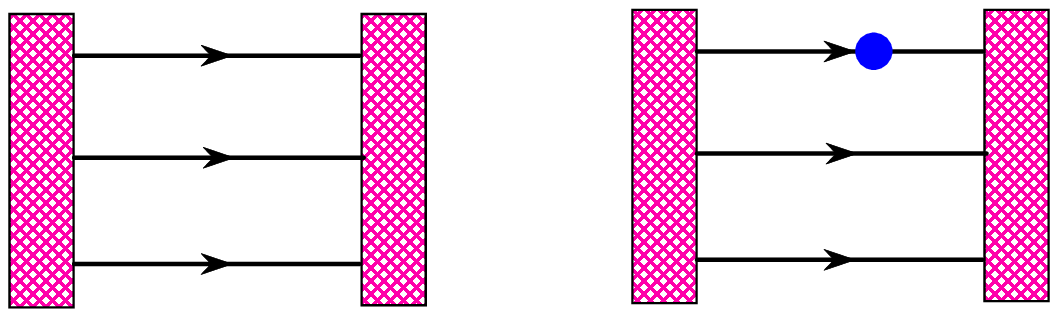}
\caption{Graphs showing the normalization of a 3-quark component of
a baryon (left) and the matrix element of a local operator denoted
by a circle (right).} \label{fig:6}
\end{minipage}
\hfil
\begin{minipage}[t]{.40\textwidth}
\includegraphics[width=\textwidth]{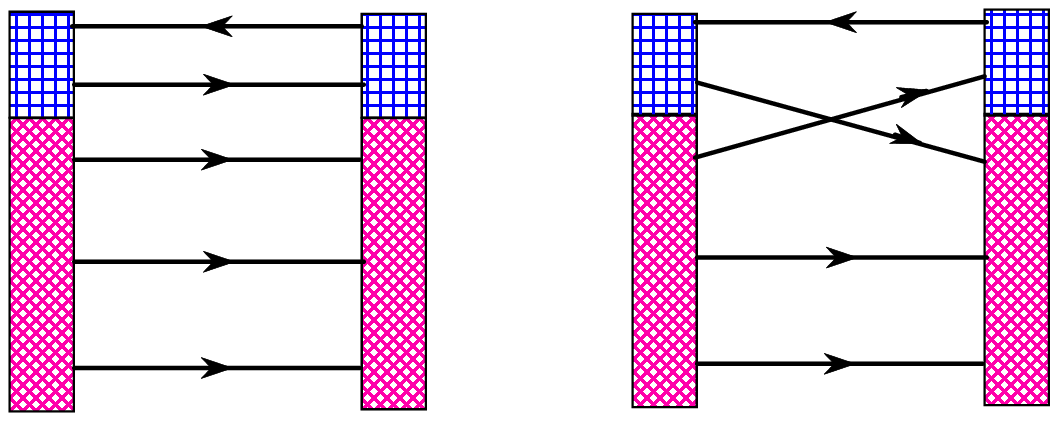}
\caption{Direct (left) and exchange (right) contributions to the
normalization of the 5-quark component of a baryon. The upper
rectangles denote $Q\bar Q$ pairs.} \label{fig:7}
\end{minipage}
\end{figure}

In the matrix elements between the $3Q$ components the $b^\d b$ part
of the current is passive as there are no antiquarks. The $a^\d a$
part is a sum over colors. As in the normalization, one gets the
factor $6\cdot 6$ from all contractions. Let it be the third quark
whose charge is measured: there is a factor of 3 from three quarks
to which the charge operator can be applied, see Fig.~6. Denoting
for short $\int\!(dp_{1-3})$ the integrals over momenta with the
conservation $\delta$-functions as in \eq{N31} we arrive at the
following expression for the matrix element of the vector charge:
\beqa\nn
V^{(3)}(1\to 2)&=&\frac{(6\!\cdot\!6\!\cdot\!3)\!}{2}\,\delta^k_l\,T(1)^{f_1f_2f_3}_{j_1j_2j_3,k}\,
T(2)^{l_1l_2l_3,l}_{f_1f_2g_3}\int\!(dp_{1-3})\\
&&\cdot
\left[F^{j_1\sigma_1}(p_1)F^{j_2\sigma_2}(p_2)F^{j_3\sigma_3}(p_3)\right]
\,\left[F^\d_{l_1\sigma_1}(p_1)F^\d_{l_2\sigma_2}(p_2)F^\d_{l_3\tau_3}(p_3)\right]\,
\left[\delta^{\tau_3}_{\sigma_3}J^{g_3}_{f_3}\right].
\la{V31}\eeqa
One can easily check using \eq{DpDn} that, say, for the $p\to
n\rho^+$ transition, the above vector charge gives exactly the same
expression as for the normalization \ur{N31}. Therefore, the $g_V$
of this transition is unity, as it should be for the conserved
vector current.

We consider here for simplicity only matrix elements of operators
with zero momentum transfer. If it is non-zero, the generalization is obvious:
one has to change the momentum of one of the quarks on which the operator acts,
by the corresponding momentum transfer, and leave the rest quarks momenta
unaltered.

For the axial transition, one replaces averaging over baryon spin by
$\half(-\sigma_3)^k_l$, and the axial charge operator is now
$(-\sigma_3)^{\tau_3}_{\sigma_3}$ instead of
$\delta^{\tau_3}_{\sigma_3}$, see \eq{vac}. All the rest is the same
as in \eq{V31}:
\beqa\nn
A^{(3)}(1\to 2)&=&\frac{(6\!\cdot\!6\!\cdot\!3)\!}{2}\,(-\sigma_3)^k_l\,T(1)^{f_1f_2f_3}_{j_1j_2j_3,k}\,
T(2)^{l_1l_2l_3,l}_{f_1f_2g_3}\int\!(dp_{1-3})\\
&&\!\cdot\!\left[F^{j_1\sigma_1}(p_1)F^{j_2\sigma_2}(p_2)F^{j_3\sigma_3}(p_3)\right]
\left[F^\d_{l_1\sigma_1}(p_1)F^\d_{l_2\sigma_2}(p_2)F^\d_{l_3\tau_3}(p_3)\right]\!
\left[(\!-\sigma_3)^{\tau_3}_{\sigma_3}J^{g_3}_{f_3}\right]\!.
\la{A31}\eeqa
The result, however, is now different as the axial
charge is not conserved. For example, for the $p\to n\pi^+$
transition one gets the expression identical to that for the
normalization but with the factor 5/3. It means that we have
obtained in the non-relativistic limit for the $3Q$ component of the
nucleon $g_{A}^{(3)}(N)=5/3$. It is the well-known result of the
non-relativistic quark model. However, it is modified by the
relativistic corrections to the valence quark wave functions
\urs{Flev_IMF}{Fsea_IMF} and by the $5Q$ component of the nucleon.

\section{Five quarks: normalization, vector and axial charges}

Already in the normalization of the $5Q$ Fock component of a baryon
there are two types of contributions: direct and exchange ones, see
Fig.~7. In the former, one contracts $a^\d$ from the pair wave
function with an $a$ in the conjugate pair, and all the valence
operators are contracted with each other. There are 6 such
possibilities, and the contraction in color gives a factor $3\cdot
6$, all in all 108. In the exchange contributions, one contracts
$a^\d$ from the pair with one of the three $a$'s from the valence
level. Further on, $a$ from the conjugate pair is contracted with
one of the three $a^\d$'s from the valence level. There are 18 such
possibilities but the contraction in color gives now only a factor
of 6. Therefore for the exchange contractions we also get a factor
of 108 but with an overall negative sign as one has to anticommute
fermion operators to get the exchange terms. As a result we obtain
the following general expression for the normalization of the $5Q$
Fock component:
\beqa\nn
{\cal N}^{(5)}(B)&=&\frac{108}{2}\,
\delta^k_l\,T(B)^{f_1f_2f_3f_4,j_5}_{j_1j_2j_3j_4,f_5,k}\,
T(B)^{l_1l_2l_3l_4,f_5,l}_{f_1f_2g_3g_4,l_5}\int\!(dp_{1-5})\\
\nn &\cdot
&F^{j_1\sigma_1}(p_1)F^{j_2\sigma_2}(p_2)F^{j_3\sigma_3}(p_3)\,
W^{j_4\sigma_4}_{j_5\sigma_5}(p_4,p_5)\,F^\dagger_{l_1\sigma_1}(p_1)
F^\dagger_{l_2\sigma_2}(p_2)\\
\la{N51} &\cdot
&\left[F^\dagger_{l_3\sigma_3}(p_3)W^{l_5\sigma_5}_{\!c\,l_4\sigma_4}(p_4,p_5)\,
\delta^{g_3}_{f_3}\delta^{g_4}_{f_4}
-F^\dagger_{l_3\sigma_4}(p_4)W^{l_5\sigma_5}_{\!c\,l_4\sigma_3}(p_3,p_5)
\delta_{f_4}^{g_3}\delta_{f_3}^{g_4}\right]
\eeqa
where we have denoted
\beq
\int\!(dp_{1-5})=\int\!dz_{1-5}\,\delta(1-z_1-...-z_5)
\int\frac{d^2{\bf p_{1-5,\,\perp}}}{(2\pi)^{10}}\,(2\pi)^2\,
\delta({\bf p_{1\perp}}+...+{\bf p_{5\perp}}).
\la{p15}\eeq
The flavor tensor here is the group integral projecting the $5Q$ state
onto a particular baryon, see \eq{T5}.

\begin{figure}[htb]
\begin{minipage}[t]{.40\textwidth}
\includegraphics[width=\textwidth]{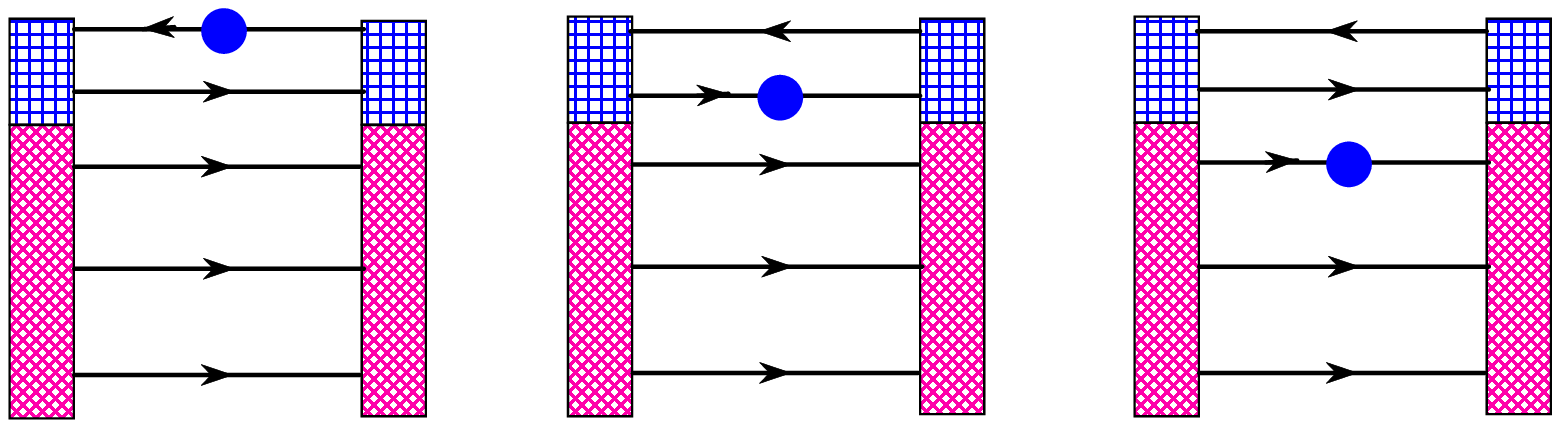}
\caption{Direct contributions to the matrix element of an operator,
in the 5-quark component of a baryon. The operator is applied to the
antiquark (left), to the quark from the pair (middle) and to the
quark from the valence level (right).} \label{fig:8}
\end{minipage}
\hfil
\begin{minipage}[t]{.50\textwidth}
\includegraphics[width=\textwidth]{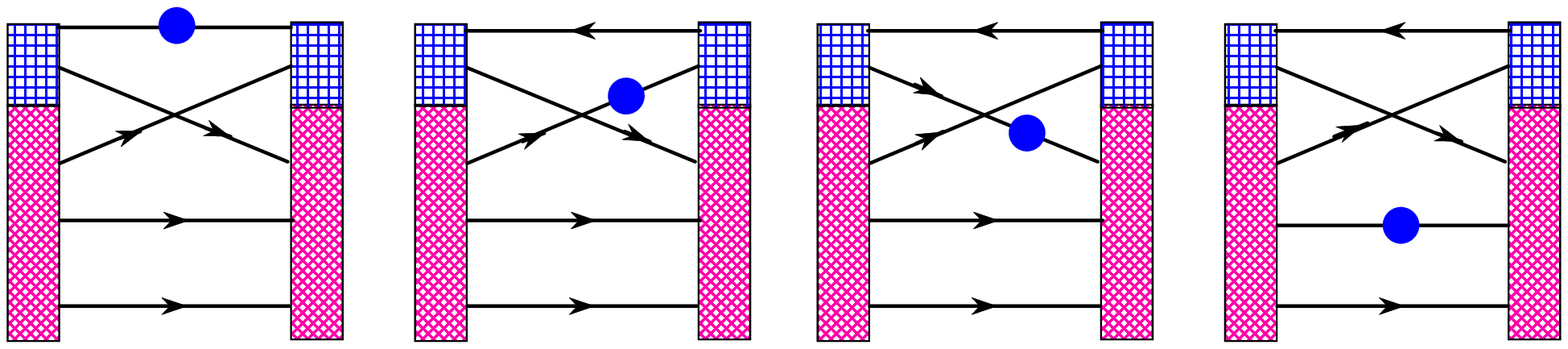}
\caption{Four types of exchange contributions to the matrix element
in the 5-quark component of a baryon.} \label{fig:9}
\end{minipage}
\end{figure}

The ratio of the normalization factors ${\cal N}^{(5)}/{\cal N}^{(3)}$
gives the probability to find a $5Q$ component in a mainly $3Q$
baryon. It depends on the mean field inside a baryon through the
pair wave function $W$ (and is quadratic in the mean field), and on
the particular baryon through its spin-flavor content $T$.

For the vector and axial transitions there are three basic
contributions: one when the charge of the antiquark is measured, the
second when the charge operator acts on the quark from the pair, and
the third when it acts on one of the three valence quarks. These
three types are further divided into the direct and exchange
contributions (Figs.~8,9). We write below only the direct
contributions; the exchange ones can be easily constructed from the
graphs in Fig.~9.

The vector transition:
\beqa\nn
V^{(5){\rm direct}}(1\to 2)&=&\frac{108}{2}\,\delta^k_l\,
T(1)^{f_1f_2f_3f_4,j_5}_{j_1j_2j_3j_4,f_5,k}\,
T(2)^{l_1l_2l_3l_4,g_5,l}_{f_1f_2g_3g_4,l_5}\int\!(dp_{1-5})\\
\nn &\cdot
&F^{j_1\sigma_1}(p_1)F^{j_2\sigma_2}(p_2)F^{j_3\sigma_3}(p_3)\,
W^{j_4\sigma_4}_{j_5\sigma_5}(p_4,p_5)\,F^\dagger_{l_1\sigma_1}(p_1)
F^\dagger_{l_2\sigma_2}(p_2)F^\dagger_{l_3\tau_3}(p_3)\,
W^{l_5\tau_5}_{\!c\,l_4\tau_4}(p_4,p_5)\\
\la{V51} &\cdot
&\left[-\delta_{f_3}^{g_3}\delta_{f_4}^{g_4}J_{g_5}^{f_5}\delta^{\tau_3}_{\sigma_3}
\delta^{\tau_4}_{\sigma_4}\delta^{\sigma_5}_{\tau_5}
+\delta_{f_3}^{g_3}J_{f_4}^{g_4}\delta_{g_5}^{f_5}\delta^{\tau_3}_{\sigma_3}
\delta^{\tau_4}_{\sigma_4}\delta^{\sigma_5}_{\tau_5}+
3J_{f_3}^{g_3}\delta_{f_4}^{g_4}\delta_{g_5}^{f_5}\delta^{\tau_3}_{\sigma_3}
\delta^{\tau_4}_{\sigma_4}\delta^{\sigma_5}_{\tau_5}\right].
\eeqa

The axial transition:
\beqa\nn
A^{(5){\rm direct}}(1\to 2)&=&\frac{108}{2}\,(-\sigma_3)^k_l\,
T(1)^{f_1f_2f_3f_4,j_5}_{j_1j_2j_3j_4,f_5,k}\,
T(2)^{l_1l_2l_3l_4,g_5,l}_{f_1f_2g_3g_4,l_5}\int\!(dp_{1-5})\\
\nn &\cdot
&F^{j_1\sigma_1}(p_1)F^{j_2\sigma_2}(p_2)F^{j_3\sigma_3}(p_3)\,
W^{j_4\sigma_4}_{j_5\sigma_5}(p_4,p_5)\,F^\dagger_{l_1\sigma_1}(p_1)
F^\dagger_{l_2\sigma_2}(p_2)F^\dagger_{l_3\tau_3}(p_3)\,
W^{l_5\tau_5}_{\!c\,l_4\tau_4}(p_4,p_5)\\
\la{A51} &\cdot
&\left[\delta_{f_3}^{g_3}\delta_{f_4}^{g_4}J_{g_5}^{f_5}\delta^{\tau_3}_{\sigma_3}
\delta^{\tau_4}_{\sigma_4}(\sigma_3)^{\sigma_5}_{\tau_5}
-\delta_{f_3}^{g_3}J_{f_4}^{g_4}\delta_{g_5}^{f_5}\delta^{\tau_3}_{\sigma_3}
(\sigma_3)^{\tau_4}_{\sigma_4}\delta^{\sigma_5}_{\tau_5}
-3J_{f_3}^{g_3}\delta_{f_4}^{g_4}\delta_{g_5}^{f_5}(\sigma_3)^{\tau_3}_{\sigma_3}
\delta^{\tau_4}_{\sigma_4}\delta^{\sigma_5}_{\tau_5}\right],
\eeqa
where $J^e_h$ is the flavor content of the current defined in the
previous section.

In the next sections we apply these general formulae to the calculation of
the nucleon axial charge and the $\Theta^+$ width.

\section{Five quarks: overlap integrals in the infinite momentum frame}

It takes a few minutes by {\it Mathematica} to perform the contractions
in \eqss{N51}{V51}{A51} over all flavor ($f,g$), isospin ($j,l$) and spin ($\sigma,\tau$)
indices. After all contractions are performed, one is left with
scalar integrals over longitudinal ($z$) and transverse (${\bf p_\perp}$) momenta
of the five quarks. The integrals over the relative transverse momenta in the
$\bar QQ$ pair are generally UV divergent, reflecting the divergence of the
negative-energy Dirac sea of quarks (Fig.~1). In reality, this divergence
is cut by the momentum-dependent dynamical quark mass $M(p)$, see \eq{action}.
Following Ref.~\cite{SF} where parton distributions in nucleons have been computed,
satisfying all general sum rules, we shall mimic the fall-off of $M(p)$ by
the Pauli--Villars cutoff at $M_{\rm PV}=557\,{\rm MeV}$ (this value is chosen
from the requirement that the constant $F_\pi=93\,{\rm MeV}$ is reproduced from
$M(0)=345\,{\rm MeV}$).

The pair wave function $W$ \ur{W} is determined by the Fourier transforms
of the mean chiral field $\Pi({\bf q})$ and $\Sigma({\bf q})$
\ur{SigmaPi_F}:
we find
\beqa\la{Pi1}
&&\Pi({\bf q})^i_j=i\frac{(q^a\tau^a)^i_j}{|{\bf q}|}\,\Pi(q),\qquad
\Pi(q)=\frac{4\pi}{q^2}\int_0^\infty\!dr\,\sin P(r)(qr \cos qr -\sin qr)<0,  \\
\la{Sigma1}
&&\Sigma({\bf q})^i_j=\delta^i_j\,\Sigma(q),\qquad\,
\Sigma(q)=\frac{4\pi}{|q|}\int_0^\infty\!dr\,r\,\left(\cos P(r)-1\right)\,\sin qr<0.
\eeqa
Actually ${\bf q}$ is the 3-momentum of the $\bar QQ$ pair, which in the
infinite momentum frame is
${\bf q}=({\bf p_{4\perp}}+{\bf p_{5\perp}},\,(z_4+z_5){\cal M})$.

In the ``direct" $5Q\to 5Q$ transitions \urss{N51}{V51}{A51} with zero momentum
transfer the following four scalar integrals arise from squaring \eq{W1},
corresponding to i) the full square of $\Pi({\bf q})$, ii) the square of
$\Sigma({\bf q})$, iii) the square of the third component $\Pi_3({\bf q})$,
and iv) the mixed $\Pi_3({\bf q})\Sigma({\bf q})$ term:
\beqa
\la{Kpp}
K_{\pi\pi}\!\!
&=&\frac{M^2}{2\pi}\!\!\int\!\!\frac{d^3{\bf q}}{(2\pi)^3}\,
\Phi\left(\frac{q_z}{{\cal M}},{\bf q}_\perp\right)\,
\theta(q_z)q_z\,
\Pi^2({\bf q})\!\int_0^1\!\!dy\!\int\!\frac{d^2{\bf Q}_\perp}{(2\pi)^2}
\left[\frac{{\bf Q}_\perp^2+M^2}{\left({\bf Q}_\perp^2+M^2+y(1-y){\bf q}^2\right)^2}
-(M\!\to \!M_{\rm PV})\right],\\
\la{Kss}
K_{\sigma\sigma}\!\!
&=&\frac{M^2}{2\pi}\!\!\int\!\!\frac{d^3{\bf q}}{(2\pi)^3}\,
\Phi\left(\frac{q_z}{{\cal M}},{\bf q}_\perp\right)\,
\theta(q_z)q_z\,
\Sigma^2({\bf q})\!\int_0^1\!\!dy\!\int\!\frac{d^2{\bf Q}_\perp}{(2\pi)^2}
\left[\frac{{\bf Q}_\perp^2+M^2(2y-1)^2}{\left({\bf Q}_\perp^2+M^2+y(1-y){\bf q}^2\right)^2}
-(M\!\to \!M_{\rm PV})\right],\\
\la{K33}
K_{33}\!\!
&=&\frac{M^2}{2\pi}\!\!\int\!\!\frac{d^3{\bf q}}{(2\pi)^3}\,
\Phi\left(\frac{q_z}{{\cal M}},{\bf q}_\perp\right)\,
\theta(q_z)\frac{q_z^3}{{\bf q}^2}\,
\Pi^2({\bf q})\!\int_0^1\!\!dy\!\int\!\frac{d^2{\bf Q}_\perp}{(2\pi)^2}
\left[\frac{{\bf Q}_\perp^2+M^2}{\left({\bf Q}_\perp^2+M^2+y(1-y){\bf q}^2\right)^2}
-(M\!\to \!M_{\rm PV})\right],\\
\la{K3s}
K_{3\sigma}\!\!
&=&\frac{M^2}{2\pi}\!\!\int\!\!\frac{d^3{\bf q}}{(2\pi)^3}\,
\Phi\left(\frac{q_z}{{\cal M}},{\bf q}_\perp\right)
\theta(q_z)\frac{q_z^2}{|{\bf q}|}\,
\Pi({\bf q})\Sigma({\bf q})\!\!\int_0^1\!\!\!dy\!\!\int\!\frac{d^2{\bf Q}_\perp}{(2\pi)^2}\!
\left[\!\frac{{\bf Q}_\perp^2+M^2(2y-1)}{\left({\bf Q}_\perp^2+M^2+y(1-y){\bf q}^2\right)^2}
-(M\!\to \!M_{\rm PV})\!\right]\!.
\eeqa
We have rearranged the integrals $dp_{1\!-\!5}$ such that we first integrate
over the relative momenta inside the $\bar QQ$ pair $y, {\bf Q}_\perp$ (see
\eq{zamena}) and then over the 3-momentum ${\bf q}$ of the pair as a whole.
As explained above, we regularize all integrals over the relative momenta
by the Pauli--Villars subtraction. The step function $\theta(q_z)$ ensures
that in the IMF the longitudinal momentum carried by the pair is positive.
By $\Phi(z,{\bf q}_\perp)$ we denote the probability that three valence quarks
``leave" the longitudinal fraction $z=z_4+z_5=q_z/{\cal M}$ and the transverse
momentum ${\bf q}_\perp={\bf p}_{4\perp}+{\bf p}_{5\perp}$ to the $\bar QQ$ pair:
\beqa
\la{Phi}
\Phi(z,{\bf q}_\perp)
&=&\int\frac{d^2{\bf p}_{1,2,3\,\perp}}{(2\pi)^6}dz_{1,2,3}\,
(2\pi)^2\delta({\bf p}_{1\perp}+{\bf p}_{2\perp}+{\bf p}_{3\perp}
+{\bf q}_\perp)\,\delta(1-z-z_1-z_2-z_3)\,h^2({\bf p}_1)
h^2({\bf p}_2)h^2({\bf p}_3),\\
\nn
&&{\bf p}_{1,2,3}=({\bf p}_{\perp\,1,2,3},p_{z\,1,2,3}),\qquad
p_{z\,1,2,3}=z_{1,2,3}{\cal M}-E_{\rm lev}\,.
\eeqa
In the $3Q$ components of baryons, there are no additional $\bar QQ$ pairs,
and all quantities considered in Section IX are proportional to $\Phi(0,0)$.
Since the normalization of the discrete-level wave function $h({\bf p})$ is
arbitrary, we choose it such that $\Phi(0,0)=1$.\\

Let us give a few examples how the normalization, vector and axial charges of
the $5Q$ components of baryons are expressed through the integrals
(\ref{Kpp}-\ref{K3s}) after all contractions in \eqss{N51}{V51}{A51}
are performed.\\

\noi
Nucleon normalization:
\beq
{\cal N}^{(5)}(N)=\frac{18}{5}(11 K_{\pi\pi}+23 K_{\sigma\sigma}).
\la{NN5}\eeq
For the vector charge of the $n\to p$ transition one gets exactly the same
expression, which demonstrates that the vector charge is conserved in each Fock
component separately. The vector charge of the $\Theta^+\to K^+n$ transition
turns out to be identically zero: it reflects the known fact that matrix
elements of $SU(3)$ flavor generators between different irreducible representations,
in this case between ${\bf\bar 10}$ and ${\bf 8}$, are zero; it serves as
an additional check of \eq{V51} since individual contributions in that equation
are non-zero.\\

\noi
Nucleon axial charge:
\beq
A^{(5)}(p\to\pi^+n)=\frac{6}{25}(209 K_{\pi\pi}+559 K_{\sigma\sigma}
-34 K_{33}-356 K_{3\sigma}).
\la{AN5}\eeq

\noi
$\Theta^+$ normalization:
\beq
N^{(5)}(\Theta)=\frac{36}{5}(K_{\pi\pi}+K_{\sigma\sigma}).
\la{NTheta5}\eeq

\noi
Axial charge of the $\Theta^+\to K^+n$ transition:
\beq
A^{(5)}(\Theta^+\to K^+n)=\frac{6}{5}\sqrt{\frac{3}{5}}(-7 K_{\pi\pi}-5 K_{\sigma\sigma}
+8 K_{33}+28 K_{3\sigma}).
\la{ATheta5}\eeq
Notice that the coefficients are an order of magnitude less in the $\Theta^+$
than in the nucleon case. It should be noted that we have independently derived
eqs. (\ref{NN5}-\ref{ATheta5}) in another way by applying the
charge operators directly to the five quarks and using the $SU(3)$ Clebsch--Gordan
machinery for projecting the $5Q$ states onto the baryons in question.
Since this technique is different from the one presented here, it serves as
a powerful check of the above expressions. We now proceed to evaluate them numerically.

\section{Numerical results}

For the numerical evaluation of the integrals involved in the $5Q$ matrix
elements we use the quark mass $M=345\,{\rm MeV}$, the self-consistent profile
function \ur{sc_prof}, the Pauli--Villars mass $M_{\rm PV}=557\,{\rm MeV}$, and
the baryon mass ${\cal M}=1207\,{\rm MeV}$, as it follows for the ``classical''
mass ({\it i.e.} without quantum corrections) in the mean field approximation
\cite{DPPrasz}. The self-consistent pseudoscalar $\Pi({\bf q})$ and scalar
$\Sigma({\bf q})$ fields, as given by \eqs{Pi1}{Sigma1} are plotted in Fig.~10.
The probability distribution $\Phi(z,{\bf q}_\perp)$ \ur{Phi} that the $\bar QQ$
pair carries the fraction $z$ of the baryon momentum and the transverse momentum
${\bf q}_\perp$ is plotted in Fig.~11.

\begin{figure}[htb]
\begin{minipage}[t]{.45\textwidth}
\includegraphics[width=\textwidth]{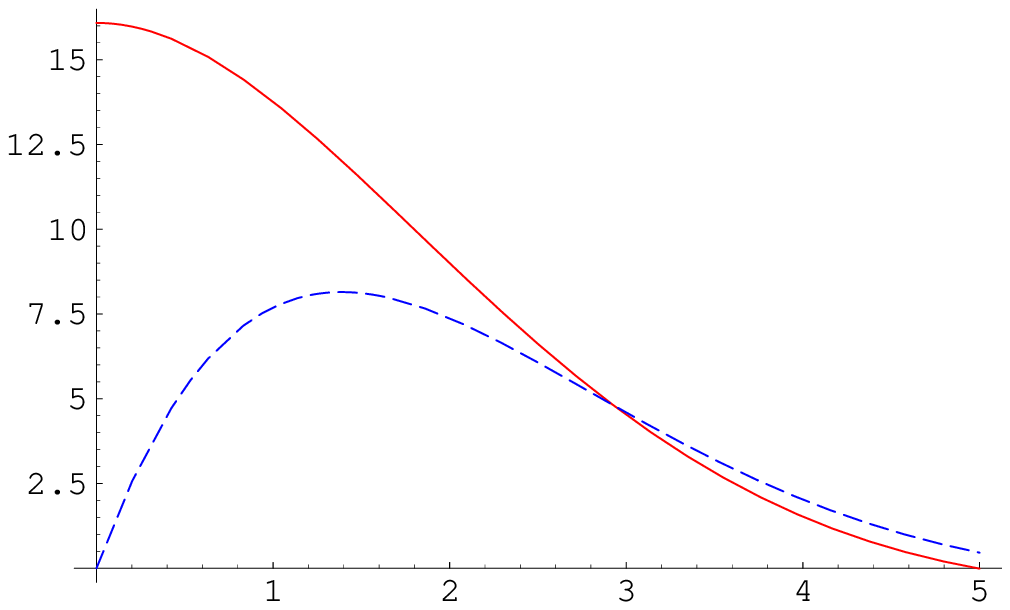}
\caption{The self-consistent pseudoscalar $-|{\bf q}|\Pi({\bf q})$ (solid)
and scalar $-|{\bf q}|\Sigma({\bf q})$ (dashed) fields in baryons in the
momentum representation. The horizontal axis is in units of $M=345\,{\rm MeV}$.}
\label{fig:10}
\end{minipage}
\hfil
\begin{minipage}[t]{.45\textwidth}
\includegraphics[width=\textwidth]{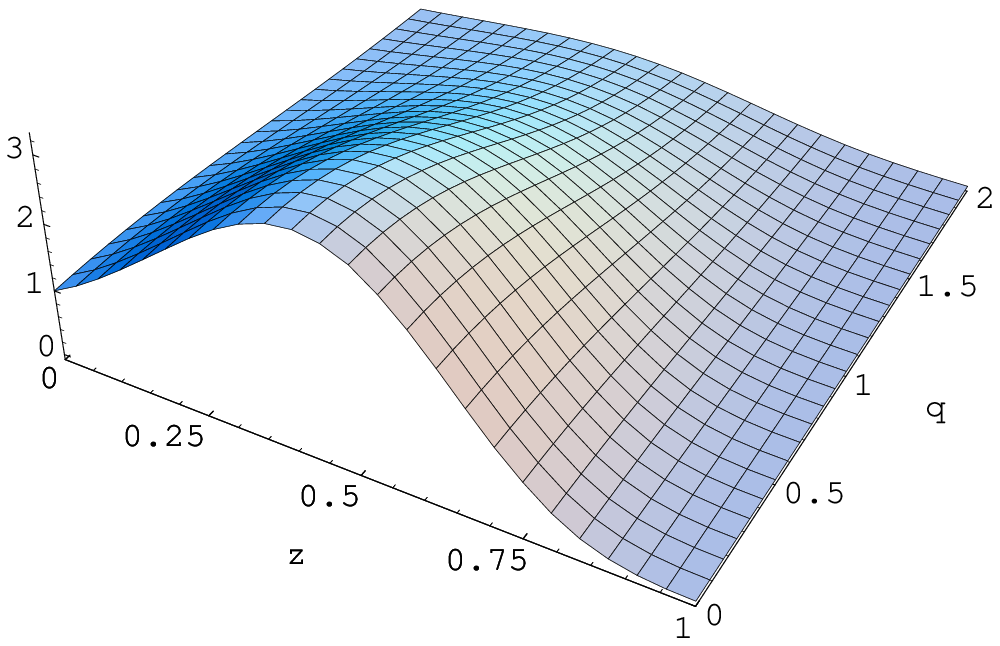}
\caption{The probability distribution $\Phi(z,{\bf q}_\perp)$ that the
$\bar QQ$ pair carries the fraction $z$ of the baryon momentum and the
transverse momentum ${\bf q}_\perp$ plotted in units of $M=345\,{\rm MeV}$.}
\label{fig:11}
\end{minipage}
\end{figure}

With these functions, the numerical evaluation of the integrals
(\ref{Kpp}-\ref{K3s}) yields
\beq
K_{\pi\pi}=0.0623,\qquad K_{\sigma\sigma}=0.0284,\qquad K_{33}=0.0372,\qquad K_{3\sigma}=0.0333.
\la{Knum}\eeq
Putting these values into eqs.(\ref{NN5}-\ref{ATheta5}) we obtain\\

\noindent
Nucleon $5Q$ normalization:
\beq
{\cal N}^{(5)}(N)=4.813.
\la{NN51}\eeq
Nucleon $5Q$ axial charge:
\beq
A^{(5)}(p\to\pi^+n)=3.779.
\la{AN51}\eeq
$\Theta^+$ $5Q$ normalization:
\beq
N^{(5)}(\Theta)=0.652.
\la{NTheta51}\eeq
$\Theta^+$ $5Q$ axial charge:
\beq
A^{(5)}(\Theta^+\to K^+n)=0.607.
\la{ATheta51}\eeq

One has to add the $3Q$ nucleon normalization computed
from \eq{N31}
\beq
{\cal N}^{(3)}(N)=9\Phi(0,0)=9
\la{NN31}\eeq
and the $3Q$ nucleon axial charge computed from \eq{A31}
\beq
A^{(3)}(p\to\pi^+n)=15\Phi(0,0)=15,
\la{AN31}\eeq
from where it follows that in the non-relativistic $3Q$ approximation the
nucleon axial charge is
\beq
g_A^{(3)}(N)=\frac{A^{(3)}(p\to\pi^+n)}{{\cal N}^{(3)}(N)}
=\frac{5}{3}\approx 1.67
\la{gAN3}\eeq
which is the well-known result of the non-relativistic quark model.

In the $5Q$ approximation, the nucleon axial charge is
\beq
g_A^{(5)}(N)=\frac{A^{(3)}(p\to\pi^+n)+A^{(5)}(p\to\pi^+n)}
{{\cal N}^{(3)}(N)+{\cal N}^{(5)}(N)}\approx 1.36
\la{gAN5}\eeq
which brings it closer to the experimental value $g_A(N)=1.27$.
The account for any number of pairs and for relativistic corrections
in the $1/N_c$ expansion brings $g_A$ very close to the experimental
value~\cite{corrNc1}.

We note that the ratio of the $5Q$ to the $3Q$ normalization in the nucleon is
\beq
\frac{{\cal N}^{(5)}(N)}{{\cal N}^{(3)}(N)}=0.535\approx 50\%.
\la{N5toN3}\eeq
On the one hand, it means that the $5Q$ Fock component of the nucleon is
quite substantial but on the other hand it implies
that antiquarks carry roughly only
$$\frac{0\cdot 1+\frac{1}{5}\cdot\half}{1+\half}\approx 7\%$$
of the nucleon momentum, assuming the antiquark carries $1/5$ of the momentum
in the $5Q$ component~\cite{footnote2}.
We have not evaluated the $7Q...$ normalization in the nucleon (which would follow
from expanding the coherent exponent in \eq{Psi} to higher orders) but expect
that higher Fock components are suppressed, roughly, by factorials following
from the expansion of the exponent. At large $N_c$, however, there would be on the
average ${\cal O}(N_c)$ $\bar QQ$ pairs in the nucleon.

Turning to the axial constant of the $\Theta^+\to KN$ transition we obtain
\beq
g_A(\Theta\to KN)=\frac{A^{(5)}(\Theta^+\to K^+n)}
{\sqrt{N^{(5)}(\Theta)}\,\sqrt{{\cal N}^{(3)}(N)+{\cal N}^{(5)}(N)}}
=0.202
\la{gATheta}\eeq
being substantially less than the nucleon axial charge computed in the same
approximation. The quantity is similar in spirit (and magnitude) not to the
nucleon axial coupling itself but to its {\em change} as due to the $5Q$
component in the nucleon, $g_A^{(3)}(N)-g_A^{(5)}(N)=0.31$. It is additionally
suppressed by the $SU(3)$ group factors for the ${\bf\overline{10}}\to{\bf 8}$
transition.

Assuming the approximate $SU(3)$ chiral symmetry (which was the base
for using the $\Theta^+$ wave function \ur{Theta1} in the first place) one can
get the $\Theta^+\to KN$ pseudoscalar coupling from the generalized Goldberger--Treiman
relation
\beq
g_{\Theta KN}=\frac{g_A(\Theta\to KN)(M_\Theta+M_N)}{2F_K}=2.24
\la{gTheta}\eeq
where we use $M_\Theta=1530\,{\rm MeV},\,M_N=940\,{\rm MeV},\,F_K=1.2F_\pi=112\,{\rm MeV}$.
Knowing the transition pseudoscalar constant one can evaluate the $\Theta^+$
width from the general expression for the $\half^+$ hyperon decay~\cite{Okun}
\beq
\Gamma_{\Theta}=2\cdot\frac{g_{\Theta KN}^2|{\bf p}|}{8\pi}\,
\frac{(M_\Theta-M_N)^2-m_K^2}{M_\Theta^2}=4.44\,{\rm MeV}
\la{GammaTheta}\eeq
where $|{\bf p}|=\sqrt{(M_\Theta^2-M_N^2-m_K^2)^2-4M_N^2m_K^2}/2M_\Theta=254\,{\rm MeV}$
is the kaon momentum in the decay ($m_K=495\,{\rm MeV}$), and we have put the
factor 2 to account for the equal-probability $K^+n$ and $K^0p$ decays.

\section{Theoretical uncertainties}

Unfortunately, in baryon physics we deal with a truly strong interaction case,
meaning that all dimensionless quantities are generally speaking of the order
of unity. There is no really small algebraic parameter in sight that would
allow some kind of perturbative expansion. We have argued in the Introduction
that $1/N_c$ can be considered as a formal small parameter justifying the use
of the Relativistic Mean Field Approximation. However, it is definitely not small
enough when it comes to ``kinematical'' factors related to the rotational states
of the mean-field baryons. Therefore, we treat the octet, decuplet and
antidecuplet baryons as it should be at $N_c=3$, rather than dealing with the
large-$N_c$ prototypes of those multiplets. We expect that the accuracy of this
mixed logic is at the level of $1/(2\pi N_c)\sim 10\%$.

Another source of the uncertainty is the present lack of knowledge of the
exact low-energy effective action \ur{action}, in particular of the
exact dynamical quark mass $M(p)$. We have mimicked the fall-off of this
function at large momenta by introducing the Pauli--Villars cutoff such that
the $F_\pi$ constant and the chiral condensate $<\!\bar qq\!>$ are reproduced.
From the experience in calculating various observables in the Chiral
Quark Soliton Model~\cite{DP-CQSM} we estimate the ensuing error as
$\sim 15\%$. Thus, the resulting accuracy of the Relativistic Mean Field
Approximation with exact account for the rotational wave functions of
baryons, is expected to be about $20\%$, and this is indeed the typical
accuracy with which formfactors, magnetic moments, parton distributions etc.
have been computed in the model; in many cases the accuracy is actually much
better but we quote here the pessimistically expected accuracy.

When dealing with hyperons containing strange quarks, one has to decide how
does one treat the mass $m_s$. Theoretically speaking, there are two
small parameters, $1/N_c$ and $m_s/\Lambda$ where $\Lambda$ is the
characteristic scale of the strong interactions. Before choosing a calculational
scheme one has to decide which of the two parameters is ``smaller". One
observes that the mass splittings in the baryon octet and decuplet are
${\cal O}(m_sN_c)$ and are somewhat less than the splitting between octet
and decuplet centers, which is ${\cal O}(\Lambda/N_c)$. Also, the
Gell-Mann--Okubo relations are satisfied to the 0.5\% accuracy, which can
be algebraically written as ${\cal O}(m_s^2/\Lambda^2)$. It indicates
that the former parameter is larger than the latter, moreover it is not
unreasonable to say that the strange quark mass is very small, $m_s={\cal O}(\Lambda/N_c^2)$.
In practical terms it means that in baryons, $m_s$ can be treated as a
perturbation in most cases. In this paper, we have actually used the chiral
limit, $m_s=0$, {\it i.e.} the zeroth order of that perturbation series.
Computing first-order corrections in $m_s$ to the observables does not
cause serious difficulties, see {\it e.g.} Refs.~\cite{hyperons,DPP97},
but we have not done it here. The penalty is expected at the $20\%$ level.

Within the Relativistic Mean Field Approximation, there arises a new important
dimensionless parameter, namely $\ee=E_{\rm lev}/M$ where $M=M(0)$ is the
dynamical quark mass at zero virtuality and $E_{\rm lev}$ is the quark
bound-state level generated by the self-consistent chiral field, see \eq{level}.
If $\ee\approx 1$, the valence quarks in baryons are non-relativistic,
the upper $s$-wave Dirac component of their wave function $h(r)$ is much
larger than the lower $p$-wave component $j(r)$, and the number of additional
$\bar QQ$ pairs in baryons tends to zero. In this limit the $\Theta^+$
width goes to zero~\cite{DPP97}, which can be also seen from the equations
of the previous section, in particular from \eq{gATheta}: the numerator
in that equation ($A^{(5)}$) is proportional to the number of the $\bar QQ$
pairs while the denominator ($\sqrt{N^{(5)}(\Theta)}$) is proportional
to its square root. Consequently, the width {\bf $\Gamma_\Theta$ is proportional
to the number of $\bar QQ$ pairs in ordinary baryons} and vanishes in
the non-relativistic limit $\ee\to 1$.

Actually in our estimates in Section XII, we have systematically used
the first-order perturbation theory in the ``relativism'' of valence quarks
or, mathematically speaking, in $1-\ee$. Namely, we have
\begin{itemize}
\item ignored the lower component of the valence wave function $j(r)$
\item ignored the distortion of the valence wave function by the sea,
\eq{Fsea_IMF}
\item used the approximate expression for the pair wave function \ur{W}
\item computed the direct but neglected the exchange diagrams when evaluating
the $5Q$ normalization and transition matrix elements, shown in Fig.~7 and 9
\item neglected the $7Q,9Q...$ components in baryons.
\end{itemize}
It is difficult to evaluate the errors of these approximations before
the neglected corrections are computed (which is surely feasible as all
corrections are well defined above, but it has not been done). Unfortunately,
the uncertainty associated with this non-relativistic approximation
is expected to be large since the actual expansion parameter $1-\ee=0.42$
is poor. Another sign that the nucleon is in fact a relativistic system
comes from the $50\%$ ratio of the $5Q$ to the $3Q$ normalization.
Treating the relativistic system in the first order in the ``relativism",
is undoubtedly the main source of the uncertainty in our numerical estimates.

Assuming that the uncertainties mentioned above are ``statistically
independent", we estimate the error in computing the transition
coupling $g_{\Theta KN}$ as
$$\sqrt{0.2^2+0.2^2+0.42^2}=0.5$$
implying a 100\% error in the width.

To get a feeling of the accuracy of our estimates, we have redone the
calculations of Section XII replacing the probability distribution
$\Phi(z,{\bf q}_\perp)$ introduced in Section XI by a flat one. This is
a legitimate assumption within the Mean Field Approximation as it corresponds
to ignoring the restriction following from the quark momentum conservation.
We remind the reader that we have used the value of the baryon mass
${\cal M}=1207\,{\rm MeV}$ instead of, say, $940\,{\rm MeV}$: the difference
is believed to be partially due to adding the momentum conservation
correction to the Mean Field result~\cite{Goeke-momentum}. Therefore,
it may seem to be more logical to ignore the quark momentum conservation
systematically throughout the calculations.

With this assumption, the evaluation of the $5Q$ matrix elements
(\ref{Kpp}-\ref{K3s}) is very easy and we obtain, instead of \eq{Knum},
\beq
K_{\pi\pi}=0.0428,\qquad K_{\sigma\sigma}=0.0235,\qquad K_{33}
=0.0214,\qquad K_{3\sigma}=0.0226.
\la{Knum0}\eeq
These numbers lead, via eqs. (\ref{NN5}-\ref{ATheta5}), to the following
values of the physical quantities:
\beq
\frac{{\cal N}^{(5)}(N)}{{\cal N}^{(3)}(N)}=0.405,\qquad
g_A^{(5)}(N)=1.44,
\la{N0}\eeq
which are not qualitatively different from the estimates
\urs{gAN5}{N5toN3}. However, the $\Theta^+$ width appears to be quite
sensitive to the change:
\beq
g_A(\Theta\to KN)=0.146,\qquad \Gamma_\Theta=2.32\,{\rm MeV},
\la{GammaTheta0}\eeq
the width turning out nearly twice smaller than that of Section XII.
It gives the idea of the accuracy of our estimate.

Probably the worse error in our estimate of the $\Theta^+$ width arises
from neglecting the exchange diagrams in matrix elements, see Figs.~7 and 9.
As a rule, their account in processes involving fermions reduces matrix
elements. It should be also noticed that the mass difference between the $\Theta^+$
and the nucleon is not small whereas we have estimated the transition
amplitude at zero momentum transfer. One would hence expect that there is
an additional formfactor-like reduction of the $\Theta^+\to KN$ transition
amplitude.

Therefore, one can well imagine that the $\Theta^+$ width \ur{GammaTheta0}
is further reduced, maybe even below the $1\,{\rm MeV}$ value. We do not
think that taking into account the $7Q...$ components in the transition matrix
elements will seriously alter the $5Q$ estimates.

Pinning down the $\Theta^+$ width even inside a wide 50\% error margin
requires much more work than presented here. Nevertheless, the estimate
that $\Gamma_\Theta$ is in a few MeV range seems to be safe. It follows
from the relative suppression of $\bar QQ$ pairs in the nucleon, and
from the $SU(3)$ group suppression in the $\Theta^+\to KN$ transition.

\section{Conclusions}

Ordinary baryons are {\em not} made of three quarks only but have a substantial
component with additional $\bar QQ$ pairs. For some observables, additional
$\bar QQ$ pairs change the naive $3Q$ results by only 20\% (like in the case
of the nucleon axial constant) but for some other observables they change the
naive result by a factor of $3\!-\!4$ (as in the case of the spin carried by quarks
or the nucleon $\sigma$ term). Hence it is imperative to learn how to work with
higher Fock components in baryons.

It is imperative not only for practical but for simple theoretical reasons.
Assuming there are just three quarks in a baryon and wishing to write down
their wave function, one realizes that one cannot ``measure'' (and hence
mathematically describe) the quark position to an accuracy better than the
Compton wave length of a pion ($1.4\,{\rm fm}$), since by uncertainty principle
one then produces a pion or an additional $\bar QQ$ pair. Since the baryon size
is $1\,{\rm fm}$, there is literally no room for the just-three-quarks description
of a baryon. The uncertainty principle demands that baryons should be described
as containing an indefinite number of $\bar QQ$ pairs. The only question is
quantitative: how many are there $\bar QQ$ pairs, and what are their wave
functions~\cite{footnote3}.

Moving to this uncharted territory, one has to satisfy certain general conditions
as the relativistic invariance (since pair production is a relativistic effect)
and the completeness of states, needed to guarantee that parton distributions,
including antiquarks, are positive-definite and are subject to sum rules following
from the conservation laws for the baryon charge, axial current, etc.
Relativistic invariance and the completeness of states can be achieved only
in a relativistic quantum field theory. A field-theoretic model of baryons, which
takes into account the infinite number of degrees of freedom and in which these
general conditions are automatically met, is the Chiral Quark Soliton Model~\cite{DP-CQSM},
an alias for the Relativistic Mean Field Approximation.

Using this model, we have presented a technique allowing to write down explicitly
the $3Q$, $5Q$, $7Q...$ wave functions of the octet, decuplet and antidecuplet
baryons. In the exotic antidecuplet the $3Q$ component is, of course, absent, but
its leading $5Q$ component is space-wise similar to the non-leading $5Q$
component of the nucleon. The technique is mathematically equivalent to the
``valence quarks plus Dirac continuum" method exploited previously, but brings the
mean field approach even closer to the language of the quark wave functions used
by many people. We have shown that the standard $SU(6)$ wave functions are easily reproduced
for the $3Q$ components of the octet and decuplet baryons, if one assumes the
non-relativistic limit. However, we have given explicit formulae for the relativistic
corrections to the $3Q$ wave function, and also for the $5Q$ wave function of the nucleon
and of the exotic $\Theta^+$. Having patience one can go further and
write down {\it e.g.} the 19-quark component of the proton or the 7-quark component
of the exotic $\Xi^{--}$.

It is important that the $\bar QQ$ pair in the $5Q$ Fock component of any baryon,
be it the nucleon or the $\Theta^+$, is added in the form of a chiral field,
which costs little energy. This is the reason why the $5Q$ component in the nucleon
turns out to be substantial, and why the exotic $\Theta^+$ baryon whose Fock decomposition
starts from the $5Q$ component, is expected to be light. The energy penalty
for making a pentaquark would be exactly zero in the chiral limit and were
baryons infinitely large. In reality, to make {\it e.g.} the $\Theta^+$ from the nucleon,
one has to create a quasi-Goldstone K-meson and to confine it inside the baryon
of the size $\geq 1/M$. It costs roughly
\beq
m(\Theta)-m(N)\approx \sqrt{m_K^2+{\bf p}^2}\leq \sqrt{495^2+345^2}=603\,{\rm MeV}.
\la{cost}\eeq
Therefore, one should expect the exotic $\Theta^+$ around 1540 MeV.
The existence of the lightest degree of freedom in QCD, namely the pseudo-Goldstone
fields, is ignored in the non-relativistic constituent quark models, which
leads to the overestimate of the $\Theta^+$ mass by typically $500\,{\rm MeV}$~\cite{D04-Osaka}.

Having presented the general formalism for computing observables for the $3Q$ as
well as for higher Fock components, we have applied it to several cases of
immediate interest. We have estimated the normalization of the $5Q$ component
of the nucleon as about 50\% of the $3Q$ component, meaning that about $1/3$ of
the time the nucleon is ``made of" five quarks. We have also shown that the account
for the $5Q$ component in the nucleon moves its axial charge from the naive
non-relativistic value of $5/3$ much closer to the experimental value.

Another case of interest is the width of the exotic $\Theta^+$ baryon:
if it exists, why is it so narrow? The best direct experimental
limit is $\Gamma_\Theta<9\,{\rm MeV}$~\cite{Dolgolenko}, however indirect
estimates~\cite{narrow} indicate that the width can be as small as $1\,{\rm Mev}$
or even less. Such a narrow width for a strongly decaying baryon some $100\,{\rm MeV}$
above the threshold, is the main surprise about the $\Theta^+$. Since the
original narrow-width estimate $\Gamma_\Theta<15\,{\rm MeV}$ \cite{DPP97}
(or, to be more precise, $3.6<\Gamma_\Theta<11.4\,{\rm MeV}$ \cite{DPP04})
based on the Chiral Quark Soliton Model, we have made here the first estimate
of the axial constant for the $\Theta^+\to KN$ transition, based on the direct
computation of the $5Q$ matrix element within the same logic. We have shown
that the $\Theta^+$ width is proportional to the number of $\bar QQ$ pairs
{\em in nucleons} and is thus naturally suppressed, as compared to the expected
widths of baryons with the dominant $3Q$ component. Assuming the $SU(3)$ symmetry,
the $\Theta^+$ width is additionally suppressed by the $SU(3)$ Clebsch--Gordan factors.

In this first direct estimate using the $5Q$ wave functions of the $\Theta^+$
and of the nucleon, we have made several approximations summarized in Section XIII.
The worse approximations can be eliminated by further work outlined in the
paper but at the moment they lead to a large theoretical uncertainty in the
$\Theta^+$ width. Depending on the way we impose the approximation, we obtain
$\Gamma_\Theta\approx 2-4\,{\rm MeV}$, with a high probability that it is
further reduced by taking into account the quark exchange processes in the
$\Theta^+\to KN$ transition, and the formfactor-like suppression in this finite
momentum transfer decay (both of which we neglected). Therefore, the $\Theta^+$
width of a few MeV appears naturally within the Relativistic Mean Field Approximation,
without any parameter fixing.

We believe that the presented formalism has a broad field of applications,
apart from exotic baryons. One kind of applications has been already started
in Ref.~\cite{PP-IMF} and involves exclusive processes, nucleon distribution
amplitudes, parton distributions for a fixed number of quarks, and the like.
Another kind of applications is for low energies. One can compute any type of
transition amplitudes between various Fock components of baryons, including
the relativistic effects, the effects of the $SU(3)$ symmetry violation,
mixing of multiplets, and so on. \\

{\bf Acknowledgements}\\

V.P. is grateful to NORDITA for kind hospitality during his visit
in September--October 2003 when this work has been started. The work of
V.P. has been supported in part by the Russian Government grant 1124.2003.2.


\appendix
\section{Parametrization of $SU(N)$ matrices}

In this Appendix we construct by induction a parametrization of a general
unitary unimodular $SU(N)$ matrix in terms of $N^2-1$ ``Euler angles",
and write down the invariant Haar measure over the group in terms of these angles.
The construction has been prompted by the parametrization of the $SU(3)$ group
by Mathur and Sen~\cite{MS}.

The idea is to write iteratively a general $SU(N)$ matrix as
\beq
R_N=S_NR_{N-1}
\la{RN}\eeq
where $R_{N-1}$ is a general $SU(N-1)$ matrix with $(N-1)^2-1$ parameters
and $S_N$ is an $SU(N)$ matrix of a special kind with only $2N-1$ parameters
belonging to the sphere $S^{2N-1}$. It gives the full parametrization
of a general $SU(N)$ matrix with $N^2-1$ parameters. Accordingly, the
invariant integration measure over the $SU(N)$ group is presented as a product
of measures over the spheres $S^3\times S^5\times ... \times S^{2N-1}$.

One starts from the $SU(2)$ group whose parametrization as a $3d$ sphere $S^3$
is well known: one writes a general $SU(2)$ matrix as
\beq
S_2=\left(\ba{cc}e^{-i\alpha_{11}}\cos\phi_1 & e^{i\alpha_{12}}\sin\phi_1 \\
-e^{-i\alpha_{12}}\sin\phi_1 & e^{i\alpha_{11}}\cos\phi_1 \ea\right)
\la{S2}\eeq
where the last column in $S_2$ can be viewed as a $2d$ complex vector $v_2=(z^1,z^2)$
normalized as $|z^1|^2+|z^2|^2=1$, which defines an $S^3$ sphere. The first
column is the orthogonal vector $v_1^i=\ee^{ij}\bar v_{2j}$. The group measure can be
written as an integral over the $S^3$ sphere,
\beq
\la{Haar21}
\frac{1}{\pi^2}\int dz^1d\bar z^1dz^2d\bar z^2\,\delta(|z^1|^2+|z^2|^2-1),
\eeq
or, explicitly in terms of three angles, as
\beq
\la{Haar2}
\frac{1}{2\pi^2}\int_0^{\frac{\pi}{2}}d\phi_1\,\sin\phi_1\cos\phi_1
\int_0^{2\pi}d\alpha_{11}\int_0^{2\pi}d\alpha_{12}
\qquad(=1).
\eeq

To construct a general $SU(3)$ matrix using the recipe \ur{RN} we first make
a $3\times 3$ matrix $R_2$ putting $S_2$, say, in its left upper corner,
\beq
R_2=\left(\ba{cc} S_2 & \ba{c} 0\\ 0\ea\\ \ba{cc}0 &0 \ea & 1 \ea\right)\,,
\la{R2}\eeq
and define
\beq
R_3=S_3R_2,\qquad
S_3=\left(\ba{ccc}e^{i\alpha_{23}}\cos\theta & 0 & e^{i\alpha_{23}}\sin\theta
\\
-e^{i\alpha_{22}}\sin\theta\sin\phi_2 &
e^{-i\alpha_{21}-i\alpha_{23}}\cos\phi_2 &
e^{i\alpha_{22}}\cos\theta\sin\phi_2 \\
-e^{i\alpha_{21}}\sin\theta\cos\phi_2 &
-e^{-i\alpha_{22}-i\alpha_{23}}\sin\phi_2 &
e^{i\alpha_{21}}\cos\theta\cos\phi_2\ea\right).
\la{R3}\eeq
The last column in $S_3$ can be viewed as a $3d$ complex
vector $v_3=(z^1,z^2,z^3)$ normalized to $|z^1|^2+|z^2|^2+|z^3|^2=1$,
which defines an $S^5$ sphere. The three columns are constructed as
(complexified) orts in spherical coordinates:
$v_1\sim e_r,\,v_2\sim e_\phi, v_3\sim e_\theta$.
There is of course a freedom of choosing the orts and the angles; we
use part of this freedom in such a way that $R_3={\bf 1_3}$ when all
angles are set to zero.

The measure on $S^5$ can be written as
\beq
\la{Haar31}
\frac{2}{\pi^3}\int dz^1d\bar z^1dz^2d\bar z^2dz^3d\bar z^3\,
\delta(|z^1|^2+|z^2|^2+|z^3|^2-1),
\eeq
or, explicitly in terms of five angles, as
\beq
\la{Haar3}
\frac{1}{\pi^3}\int_0^{\frac{\pi}{2}}d\theta\,\cos^3\theta
\sin\theta\int_0^{\frac{\pi}{2}}d\phi_2\,\sin\phi_2\cos\phi_2
\int_0^{2\pi}d\alpha_{21}\int_0^{2\pi}d\alpha_{22}\int_0^{2\pi}
d\alpha_{23}\qquad(=1).
\eeq
The integrations limits are chosen such that the $S^5$ sphere is covered once.

The full $SU(3)$ measure is found in the standard way: one constructs the metric
tensor
\beq
g_{mn}=\Tr\,\frac{\partial R_3}{\partial_{\beta^m}}\frac{\partial R_3^\dagger}
{\partial_{\beta^n}},\qquad \beta^m=\alpha_{11},\alpha_{12},\phi_1,
\alpha_{21},\alpha_{22},\alpha_{23},\phi_2,\theta,\quad m,n=1...8;
\la{metric_tensor}\eeq
then the $SU(3)$ measure is
\beq
\sqrt{\det g}\sim
(\sin\phi_1\cos\phi_1)\cdot(\cos^3\theta\sin\theta\sin\phi_2\cos\phi_2)
\la{factorization}\eeq
{\it i.e.} it is factorized into the product of the measures over the
spheres $S^3$ and $S^5$, see \eqs{Haar2}{Haar3}. All group integrals in Appendix B
can be performed directly using the above parametrization of the $SU(2)$ and
$SU(3)$ matrices and the above Haar measure. In fact we have checked the
results of Appendix B in this way.

The construction can be iteratively generalized to higher $SU(N)$ groups
in such a way that the group parameter space is a direct product of the
spheres $S^3\times S^5 \times\ldots S^{2N-1}$ with the total number of parameters
$\sum_{J=1}^{N-1}(2J+1)=N^2-1$, as it should be for the $SU(N)$
group. For example, a parametrization of $R\in SU(4)$ is
\bea
\la{R4}
R_4&=&S_4R_3,\\
\n
S_4\!\!&\!\!=\!\!&\!\!\left(\ba{cccc}
\cos\chi e^{i\alpha_{34}}& 0& 0& -\sin\chi e^{i\alpha_{34}}\\
\sin\chi\sin\theta_3 e^{i\alpha_{33}}&\cos\theta_3 e^{i\alpha_{33}} & 0 &
\cos\chi\sin\theta_3 e^{i\alpha_{33}}\\
\sin\chi\cos\theta_3\sin\phi_3 e^{i\alpha_{32}} &
-\sin\theta_3\sin\phi_3 e^{i\alpha_{32}} &
\cos\phi_3 e^{-i\alpha_{31}-i\alpha_{33}-i\alpha_{34}} &
\cos\chi\cos\theta_3\sin\phi_3 e^{-i\alpha_{32}}\\
\sin\chi\cos\theta_3\cos\phi_3 e^{i\alpha_{31}} &
-\sin\theta_3\cos\phi_3 e^{i\alpha_{31}} &
-\sin\phi_3 e^{-i\alpha_{32}-i\alpha_{33}-i\alpha_{34}} &
\cos\chi\cos\theta_3\cos\phi_3 e^{i\alpha_{31}}
\ea\right)\,,
\eea
and $R_3$ is the block-diagonal $4\times 4$ matrix with the general
$SU(3)$ matrix \ur{R3} in the left upper corner and unity in the right
lower corner. We thus add 7 new parameters to the previous 8 of the
$SU(3)$ parametrization. The columns of $S_4$ are complex orts in
spherical coordinates, $e_r,e_\theta,e_\phi,e_\chi$. Denoting the
last column $v_4=(z^1,z^2,z^3,z^4)$ the integration measure is that
of a sphere $S^7$:
\beq
\int\!dz^1d\bar z^1\ldots dz^4d\bar z^4\,\delta(|z^1|^2+\ldots|z^4|^2-1)
\sim \int\!\cos^5\chi\sin\chi\,\cos^3\theta_3\sin\theta_3\cos\phi_3\sin\phi_3\,
d\chi d\theta_3 d\phi_3 d\alpha_{31}d\alpha_{32}d\alpha_{33}d\alpha_{34}.
\la{Haar41}\eeq
The full $SU(4)$ measure built from the general rule \ur{metric_tensor}
(where now there are 15 angles) is factorized into the product of
measures over the $S^3,S^5$ and $S^7$ spheres.

\section{Group integrals}

In this Appendix, we give a list of group integrals used in the main text,
over the Haar measure of the $SU(N)$ group, normalized to unity,
$\int dR = 1$.

For any $SU(N)$ one has
\beq
\int dR\, R^f_i = 0,\qquad \int dR\,R^{\dagger\,i}_f = 0,\qquad
\int dR\,R^f_i\,R^{\dagger\,j}_g = \frac{1}{N}\delta^f_g\,
\delta^j_i.
\la{1+1}\eeq
For $N=2$ the following group integral is non-zero:
\beq
\int dR\,R^f_i\,R^g_j=\frac{1}{2}\epsilon^{fg}\,\epsilon_{ij}.
\la{2}\eeq
For $N>2$ this integral is zero; its analog in $SU(3)$ is
\beq
\int dR\,R^f_i\,R^g_j\,R^h_k
=\frac{1}{6}\epsilon^{fgh}\,\epsilon_{ijk}.
\la{3}\eeq
On the contrary, in $SU(2)$ it is zero.

The general method of finding integrals of several matrices $R,R^\dagger$ is as
follows. The result of an integration over the invariant measure can be only
invariant tensors which, for the $SU(N)$ group,
can be built solely from the Kronecker $\delta$ and Levi--Civita $\ee$ tensors.
One constructs the supposed tensor of a given rank as a combination of $\delta$'s
and $\ee$'s, satisfying the symmetry relations following from the integral in question.
The indefinite coefficients in the combination are then found from contracting
both sides with various $\delta$'s and $\ee$'s and thus by reducing the integral
to a previously derived one.

For any $SU(N)$ group one has
\beq
\int
dR\,R^{f_1}_{i_1}\,R^{\dagger\,j_1}_{g_1}
\,R^{f_2}_{i_2}\,R^{\dagger\,j_2}_{g_2}
= \frac{1}{N^2-1}\left[\delta^{f_1}_{g_1}\delta^{f_2}_{g_2}
\left(\delta^{i_1}_{j_1}\delta^{i_2}_{j_2}-\frac{1}{N}
\delta^{i_1}_{j_2}\delta^{i_2}_{j_1}\right)+
\delta^{f_1}_{g_2}\delta^{f_2}_{g_1}
\left(\delta^{i_1}_{j_2}\delta^{i_2}_{j_1}-\frac{1}{N}
\delta^{i_1}_{j_2}\delta^{i_2}_{j_1}\right)\right]
\la{2+2}\eeq
since its contraction with, say, $\delta^{g_1}_{f_1}$ must reduce it to
\eq{1+1}.

In $SU(2)$ there is an identity
\beq
\delta^j_{j_3}\epsilon_{j_1j_2}+
\delta^j_{j_1}\epsilon_{j_2j_3}+
\delta^j_{j_2}\epsilon_{j_3j_1}=0,
\la{id1+2} \eeq
using which one finds that the following integral is non-zero:
\beq
\int dR\,R^{f_1}_{j_1}\,R^{f_2}_{j_2}\,R^{f_3}_{j_3}\,R^{\dagger\,j}_{g}
=\frac{1}{6}\left(\delta^{f_1}_{g}\delta^{j}_{j_1}
\epsilon^{f_2f_3}\epsilon_{j_2j_3}+
\delta^{f_2}_{g}\delta^{j}_{j_2}
\epsilon^{f_3f_1}\epsilon_{j_3j_1}+
\delta^{f_3}_{g}\delta^{j}_{j_3}
\epsilon^{f_1f_2}\epsilon_{j_1j_2}\right).
\la{3+1}\eeq
In $SU(3)$ and higher groups this integral is zero. The analog of the
identity \ur{id1+2} in $SU(3)$ is (notice the signs in the cyclic permutation!)
\beq
\delta^i_{j_1}\epsilon_{j_2j_3j_4}-
\delta^i_{j_2}\epsilon_{j_3j_4j_1}+
\delta^i_{j_3}\epsilon_{j_4j_1j_2}-
\delta^i_{j_4}\epsilon_{j_1j_2j_3}=0,
\la{id1+3}
\eeq
and the analog of \eq{3+1} is
\bea
\n
&&\int dR\,R^{f_1}_{j_1}\,R^{f_2}_{j_2}\,R^{f_3}_{j_3}
\,R^{f_4}_{j_4}\,R^{\dagger\,j}_{g}\\
&=&
\frac{1}{24}\left(\delta^{f_1}_{g}\delta^{j}_{j_1}
\epsilon^{f_2f_3f_4}\epsilon_{j_2j_3j_4}+
\delta^{f_2}_{g}\delta^{j}_{j_2}
\epsilon^{f_3f_4f_1}\epsilon_{j_3j_4j_1}
+\delta^{f_3}_{g}\delta^{j}_{j_3}
\epsilon^{f_4f_1f_2}\epsilon_{j_4j_1j_2}
+\delta^{f_4}_{g}\delta^{j}_{j_4}
\epsilon^{f_1f_2f_3}\epsilon_{j_1j_2j_3}
\right).
\la{4+1}\eea
This integral arises when one projects three quarks from the bound-state level
onto the octet baryon.

To evaluate the $SU(3)$ average of six matrices, one needs the identities
\bea
\n
\ee_{i_1j_2j_3}\ee_{j_1i_2i_3}+\ee_{i_2j_2j_3}\ee_{i_1j_1i_3}+
\ee_{i_3j_2j_3}\ee_{i_1i_2j_1}
&=& \\
\n
\ee_{j_1i_1j_3}\ee_{j_2i_2i_3}+\ee_{j_1i_2j_3}\ee_{i_1j_2i_3}+
\ee_{j_1i_3j_3}\ee_{i_1i_2j_3}
&=&
\\
\ee_{j_1j_2i_1}\ee_{j_3i_2i_3}+\ee_{j_1j_2i_2}\ee_{i_1j_3i_3}+
\ee_{j_1j_2i_3}\ee_{i_1i_2j_3}
&=& \ee_{j_1j_2j_3}\ee_{i_1i_2i_3}.
\la{id3}\eea
One gets
\bea
\n
\int dR\,R^{f_1}_{j_1}\,R^{f_2}_{j_2}\,R^{f_3}_{j_3}\,
R^{h_1}_{i_1}\,R^{h_2}_{i_2}\, R^{h_3}_{i_3}\;\;
=\;\;\frac{1}{72}\!\!\!\!
&&\!\!\!\!\left(\ee^{f_1f_2f_3}\ee^{h_1h_2h_3}\ee_{j_1j_2j_3}
\ee_{i_1i_2i_3}\right.\\
\n
+\;\;\ee^{h_1f_2f_3}\ee^{f_1h_2h_3}\ee_{i_1j_2j_3}\ee_{j_1i_2i_3}
\;\;+\;\;\ee^{h_2f_2f_3}\ee^{h_1f_1h_3}\ee_{i_2j_2j_3}\ee_{i_1j_1i_3
}&+&\ee^{h_3f_2f_3}\ee^{h_1h_2f_1}\ee_{i_3j_2j_3}\ee_{i_1i_2j_1}\\
\n
+\;\;\ee^{f_1h_1f_3}\ee^{f_2h_2h_3}\ee_{j_1i_1j_3}\ee_{j_2i_2i_3}
\;\;+\;\;\ee^{f_1h_2f_3}\ee^{h_1f_2h_3}\ee_{j_1i_2j_3}\ee_{i_1j_2i_3
}&+&\ee^{f_1h_3f_3}\ee^{h_1h_2f_2}\ee_{j_1i_3j_3}\ee_{i_1i_2j_2}\\
+\;\;\ee^{f_1f_2h_1}\ee^{f_3h_2h_3}\ee_{j_1j_2i_1}\ee_{j_3i_2i_3}
\;\;+\;\;\ee^{f_1f_2h_2}\ee^{h_1f_3h_3}\ee_{j_1j_2i_2}\ee_{i_1j_3i_3
}&+&\left.\ee^{f_1f_2h_3}\ee^{h_1h_2f_3}\ee_{j_1j_2i_3}\ee_{i_1i_2j_3}\right).
\la{6}\eea

The result for the next integral is rather lengthy. We give it
for the general $SU(N)$. For abbreviation, we use the notation
\beq
\delta^{f_1}_{h_2}\,\delta^{f_2}_{h_3}\,\delta^{f_3}_{h_1}\;
\delta^{i_1}_{j_3}\,\delta^{i_2}_{j_2}\,\delta^{i_3}_{j_1}
\equiv
(231)(321),\; etc.
\la{abbrev}\eeq
One has
\bea
\n
&&\int dR\, R^{f_1}_{j_1}\,R^{f_2}_{j_2}\,R^{f_3}_{j_3}
\,
R^{\dagger\,i_1}_{h_1}\,R^{\dagger\,i_2}_{h_2}\,R^{\dagger\,i_3}_{h_
3}=\frac{1}{N(N^2-1)(N^2-4)}\\
\n
&\cdot&\left\{(N^2-2)\left[(123)(123)+(132)(132)+(321)(321)+
(213)(213)+(312)(231)+(231)(312)\right]\right.\\
\n
&-&N\left[(123)\left((132)+(321)+(213)\right)
+(132)\left((123)+(231)+(312)\right)
+(321)\left((312)+(123)+(231)\right)\right.\\
\n
&+&\left.(213)\left((231)+(312)+(123)\right)
+(312)\left((213)+(132)+(321)\right)
+(231)\left((321)+(213)+(132)\right)\right]\\
\n
&+&2\left[(123)\left((312)+(231)\right)+(132)\left((213)+(321)\right
)+(321)\left((132)+(213)\right)\right.\\
&+&\left.\left.(213)\left((321)+(132)\right)+(312)\left((123)+(312)
\right)+(231)\left((231)+(123)\right)\right]\right\}.
\la{3+3}
\eea
Apparently at $N=2$ something gets wrong. For $N=2$ there is a
formal identity following from the fact that at $N=2$ one has
$\ee^{f_1f_2f_3}\ee_{h_1h_2h_3}=0$:
\beq
(123)-(132)-(321)-(213)+(312)+(231)=0. \la{id23}\eeq
Consequently, for $SU(2)$ one obtains a shorter expression:
\bea
\n
&&\int dR\,
R^{f_1}_{j_1}\,R^{f_2}_{j_2}\,R^{f_3}_{j_3}
\,
R^{\dagger\,i_1}_{h_1}\,R^{\dagger\,i_2}_{h_2}\,R^{\dagger\,i_3}_{h_
3}\\\n
&=&\frac{1}{6}\left\{\left[(123)(123)+(132)(132)+(321)(321)+
(213)(213)+(312)(231)+(231)(312)\right]\right.
\\
\n
&-&\frac{1}{4}\left[(123)\left((132)+(321)+(213)\right)
+(132)\left((123)+(231)+(312)\right)
+(321)\left((312)+(123)+(231)\right)\right.\\
\n
&+&\left.\left.(213)\left((231)+(312)+(123)\right)
+(312)\left((213)+(132)+(321)\right)+(231)\left((321)+(213)
+(132)\right)\right]\right\}.
\eea

In case one is interested in the presence of an additional quark-antiquark
pair in an octet baryon, one has to use the group integral
\bea
\n
&&\int dR\,R^{f_1}_{j_1}\,R^{f_2}_{j_2}\,R^{f_3}_{j_3}\,(R^{f_4}_{j_4}\,
R^{\dagger\,j_5}_{f_5})\,R^{\dagger\,k}_g\,R^h_3=\frac{1}{360}
\\
\n
&\cdot &\left\{
\ee^{f_1f_2h}\ee_{j_1j_2}\left[\delta^{f_3}_g\delta^{f_4}_{f_5}\left
(4\delta^{j_5}_{j_4}\delta^k_{j_3}-\delta^{j_5}_{j_3}\delta^k_{j_4}\right)
+\delta^{f_4}_g\delta^{f_3}_{f_5}\left(4\delta^{j_5}_{j_3}\delta^k_{j_4}
-\delta^{j_5}_{j_4}\delta^k_{j_3}\right)\right]\right.\\
\n
&+&
\ee^{f_1f_3h}\ee_{j_1j_3}\left[\delta^{f_2}_g\delta^{f_4}_{f_5}
\left(4\delta^{j_5}_{j_4}\delta^k_{j_2}-\delta^{j_5}_{j_2}\delta^k_{j_4}
\right)+\delta^{f_4}_g\delta^{f_2}_{f_5}
\left(4\delta^{j_5}_{j_2}\delta^k_{j_4}-\delta^{j_5}_{j_4}\delta^k_{j_2}
\right)\right]\\
\n
&+&
\ee^{f_1f_4h}\ee_{j_1j_4}\left[\delta^{f_2}_g\delta^{f_3}_{f_5}
\left(4\delta^{j_5}_{j_3}\delta^k_{j_2}-\delta^{j_5}_{j_2}\delta^k_{j_3}
\right)+\delta^{f_3}_g\delta^{f_2}_{f_5}
\left(4\delta^{j_5}_{j_2}\delta^k_{j_3}-\delta^{j_5}_{j_3}\delta^k_{j_2}
\right)\right]\\
\n
&+&
\ee^{f_2f_3h}\ee_{j_2j_3}\left[\delta^{f_1}_g\delta^{f_4}_{f_5}
\left(4\delta^{j_5}_{j_4}\delta^k_{j_1}-\delta^{j_5}_{j_1}\delta^k_{j_4}
\right)+\delta^{f_4}_g\delta^{f_1}_{f_5}
\left(4\delta^{j_5}_{j_1}\delta^k_{j_4}-\delta^{j_5}_{j_4}\delta^k_{j_1}
\right)\right]\\
\n
&+&
\ee^{f_2f_4h}\ee_{j_2j_4}\left[\delta^{f_1}_g\delta^{f_3}_{f_5}
\left(4\delta^{j_5}_{j_3}\delta^k_{j_1}-\delta^{j_5}_{j_1}\delta^k_{j_3}
\right)+\delta^{f_3}_g\delta^{f_1}_{f_5}
\left(4\delta^{j_5}_{j_1}\delta^k_{j_3}-\delta^{j_5}_{j_3}\delta^k_{j_1}
\right)\right]\\
\la{5+2}
&+&\left.
\ee^{f_3f_4h}\ee_{j_3j_4}\left[\delta^{f_1}_g\delta^{f_2}_{f_5}
\left(4\delta^{j_5}_{j_2}\delta^k_{j_1}-\delta^{j_5}_{j_1}\delta^k_{j_2}
\right)+\delta^{f_2}_g\delta^{f_1}_{f_5}
\left(4\delta^{j_5}_{j_1}\delta^k_{j_2}-\delta^{j_5}_{j_2}\delta^k_{j_1}
\right)\right]\right\}.
\eea
This tensor defines, in particular, the five-quark wave function of the
nucleon, see \eq{n5}.

For finding the quark structure of the antidecuplet, the following
group integrals are relevant. The (conjugate) rotational wave
function of the antidecuplet is (see subsection IV.C)
\beq
\la{Dat1}
A^{*\{h_1\,h_2\,h_3\}}_k(R)=\frac{1}{3}(R^{h_1}_3R^{h_2}_3R^{h_3}_k+
R^{h_3}_3R^{h_1}_3R^{h_2}_k+R^{h_2}_3R^{h_3}_3R^{h_1}_k).
\eeq
Projecting it on three quarks and using \eq{6} we get an identical
zero because all terms in \eq{6} are antisymmetric in a pair of flavor indices
while the tensor \ur{Dat1} is symmetric. It reflects the fact that one cannot
build an antidecuplet from three quarks:
\beq
\int dR\,R^{f_1}_{j_1}\,R^{f_2}_{j_2}\,R^{f_3}_{j_3}\,
A^{*\{h_1\,h_2\,h_3\}}_k(R)=0.
\la{6A}\eeq
However, a similar group integral with an additional
quark-antiquark pair is non-zero:
\bea\n
&&\int dR\,R^{f_1}_{j_1}\,R^{f_2}_{j_2}\,R^{f_3}_{j_3}\,
\left(R^{f_4}_{j_4}\,R^{\dagger\,j_5}_{f_5}\right)\,
A^{*\{h_1\,h_2\,h_3\}}_k(R)=\frac{\delta^{j_5}_k}{1080}
\left\{\ee_{j_1j_2}\ee_{j_3j_4}\left[\delta^{h_3}_{f_5}
(\ee^{f_1f_2h_1}\ee^{f_3f_4h_2}+\ee^{f_1f_2h_2}\ee^{f_3f_4h_1})
\right.\right.\\\n
&+&\left.\delta^{h_1}_{f_5}
(\ee^{f_1f_2h_2}\ee^{f_3f_4h_3}+\ee^{f_1f_2h_3}\ee^{f_3f_4h_2})+
\delta^{h_2}_{f_5}
(\ee^{f_1f_2h_1}\ee^{f_3f_4h_3}+\ee^{f_1f_2h_3}\ee^{f_3f_4h_1})
\right]+\\
\n
&+&\ee_{j_2j_3}\ee_{j_1j_4}\left[\delta^{h_3}_{f_5}
(\ee^{f_2f_3h_1}\ee^{f_1f_4h_2}+\ee^{f_2f_3h_2}\ee^{f_1f_4h_1})+
\delta^{h_1}_{f_5}
(\ee^{f_2f_3h_2}\ee^{f_1f_4h_3}+\ee^{f_2f_3h_3}\ee^{f_1f_4h_2})
\right.\\
\n
&+&\left.
\delta^{h_2}_{f_5}
(\ee^{f_2f_3h_1}\ee^{f_1f_4h_3}+\ee^{f_2f_3h_3}\ee^{f_1f_4h_1})
\right]+\ee_{j_1j_3}\ee_{j_2j_4}\left[\delta^{h_3}_{f_5}
(\ee^{f_1f_3h_1}\ee^{f_2f_4h_2}+\ee^{f_1f_3h_2}\ee^{f_2f_4h_1})
\right.\\
\la{7+1A}
&+&\left.\left.\delta^{h_1}_{f_5}
(\ee^{f_1f_3h_2}\ee^{f_2f_4h_3}+\ee^{f_1f_3h_3}\ee^{f_2f_4h_2})+
\delta^{h_2}_{f_5}
(\ee^{f_1f_3h_1}\ee^{f_2f_4h_3}+\ee^{f_1f_3h_3}\ee^{f_2f_4h_1})
\right]\right\}.
\eea
In particular, for the $\Theta^+$ baryon being the 333-component
of the antidecuplet we have
\beq
\Theta^*_k(R)=\sqrt{30}A^{*\,333}_k(R)=\sqrt{30}R^3_3R^3_3R^3_k,\qquad
\Theta^k(R)=\sqrt{30}R^{\dagger 3}_3R^{\dagger 3}_3R^{\dagger k}_3\,.
\la{ThetaR}\eeq
The projection of five quarks onto to the $\Theta^+$ rotational wave function
\ur{ThetaR} gives the tensor
\beqa\n
T^{f_1f_2f_3f_4,j_5}_{j_1j_2j_3j_4,f_5,k}(\Theta)&=&
\int\!\!dR\,R^{f_1}_{j_1}R^{f_2}_{j_2}R^{f_3}_{j_3}
\left(R^{f_4}_{j_4}R^{\dagger\,j_5}_{f_5}\right)\Theta^*_k(R)\\
&=&\frac{\delta^3_{f_5}\,\delta^{j_5}_k\sqrt{30}}{180}
\left(\ee_{j_1j_2}\ee_{j_3j_4}\ee^{f_1f_2}\ee^{f_3f_4}+
\ee_{j_2j_3}\ee_{j_1j_4}\ee^{f_2f_3}\ee^{f_1f_4}+
\ee_{j_1j_3}\ee_{j_2j_4}\ee^{f_1f_3}\ee^{f_2f_4}\right).
\la{7+1Theta}\eeqa
This equation leads immediately to the five-quark wave function of the $\Theta^+$, see
\eqs{Theta1}{Theta2}.



\begin{thebibliography}{100}
\bibitem{footnote1}
In the baryon rest frame the $\bar QQ$ pair wave function is given
in Ref.~\cite{D04-Minn}.
\bibitem{footnote3}
In contrast to baryon physics, in atomic physics the ``electrons-only"
description is all right in a wide window between $10^{-11}$ and
$10^{-8}\,{\rm cm}$. Measuring the electron position to an accuracy better
than its Compton wave length produces $e^+e^-$ pairs, and we know that they
are there from the precision measurements of the radiative corrections.
Fortunately for the history of physics, their effect is small. In QCD with
its spontaneous chiral symmewtry breaking which makes pions very light,
we do not have this luxury as the effect of the additional $\bar QQ$
pairs is 100\%.
\end{thebibliography}
\end{document}